\documentclass[12pt,a4paper]{article}

\usepackage{graphicx}
\setkeys{Gin}{width=0.9\textwidth,height=0.7\textheight,keepaspectratio}

\renewcommand{\vec}[1]{\mbox{\boldmath $#1$}}  
\newcommand{\CN}{\mbox{cn}}  
\newcommand{\SN}{\mbox{sn}}  
\newcommand{\DN}{\mbox{dn}}  
\newcommand{\DC}{\mbox{dc}}  
\newcommand{\CD}{\mbox{cd}}  
\newcommand{\ND}{\mbox{nd}}  
\newcommand{\SD}{\mbox{sd}}  
\newcommand{\sech}{\mbox{sech}}  
\newcommand{\D}[2]{\frac{\partial #1}{\partial #2}}
\newcommand{\Dn}[3]{\frac{\partial^#1 #2}{\partial #3^#1}}
\newcommand{\TAV}[1]{\overline{#1}}
\newcommand{\SIM}{\textsc{sim}}
\newcommand{\OMIT}[1]{}
\newcommand{\Ord}[1]{{\cal O}\left(#1\right)}

\usepackage{url}
\usepackage[colour,hyperref]{ajr}

\begin{document}
\title{Initialization and the quasi-geostrophic slow manifold} 
\author{Stephen M. Cox\thanks{Department of Theoretical Mechanics, 
University of Nottingham, University Park, Nottingham NG7 2RD, 
\textsc{United Kingdom}.  
\protect\url{mailto:etzsmc@unicorn.nottingham.ac.uk}} \and A.J. 
Roberts\thanks{Department of Mathematics \& Computing, University of 
Southern Queensland, Toowoomba, Queensland 4350, \textsc{Australia}.  
\protect\url{mailto:aroberts@usq.edu.au}} }
\date{March 21, 1994}
\maketitle

\begin{abstract}
	Atmospheric dynamics span a range of time-scales. 
	The projection of
	measured data to a slow manifold, ${\cal M}$, removes fast gravity waves from
	the initial state for numerical simulations of the atmosphere.  
	We explore further the slow 
	manifold for a simple atmospheric model
	introduced by Lorenz and anticipate that our results will relevant to
	the vastly more detailed dynamics of atmospheres and oceans.
	
	Within the dynamics of the Lorenz model, we make clear the relation between a slow manifold $\cal M$ and the
	``slowest invariant manifold'' ({\SIM}), which was constructed by Lorenz
	in order to avoid the divergence of approximation schemes for $\cal M$.
	These manifolds are shown to be identical to within exponentially small
	terms, and so the {\SIM} in fact shares the asymptotic nature of $\cal M$.
	
	We also investigate the issue of
	balancing initial data in order to remove gravity waves.  This
	is a question of how to compute an ``initialized'' point on $\cal M$
	whose subsequent evolution matches that from the measured initial data
	that in general lie off $\cal M$.  
	We propose a choice based on the intuitive idea that the
	initialization procedure should not significantly alter the forecast.
	Numerical results demonstrate the utility of our initialization
	scheme.
	
	The normal form for Lorenz' atmospheric model shows clearly how to
	separate the dynamics of the different atmospheric waves.  However, its
	construction demonstrates that {\em any} initialization procedure must
	eventually alter the forecast---the time-scale of the divergence
	between the initialized and the uninitialized solutions is inevitable and is
	inversely proportional to the square of initial level of gravity-wave
	activity.
\end{abstract}

\tableofcontents

\section{Introduction}

Wave motion in the atmosphere has a wide range of periods: large-scale
motions are dominated by quasi-geostrophic Rossby waves, which have a
period of many days; faster inertial-gravity waves, with a period of up
to a few hours, are frequently negligible in the large
(Gill~\cite[p582]{Gill82}).  A significant problem for numerical
weather prediction is that an initial measured state of the atmosphere
contains noise which causes unphysical large-amplitude gravity waves to
arise in numerical solutions (Gill~\cite[p242]{Gill82}).  Consequently,
one is forced to take a very small time step in the numerical
integration (e.g.~Houghton~\cite[p169]{Houghton89}).  This is despite
the generally recognized feature that much of the troposphere and lower
stratosphere is quasi-geostrophic and so dominated by Rossby wave
activity (Lorenz~\cite{Lorenz86}).

To overcome this problem of unphysical gravity-wave activity, initial 
data must be ``balanced'', that is, adjusted so that subsequent 
gravity-wave activity is negligible (see e.g.~Houghton~\cite[p168]{Houghton89}, or Gill~\cite[p245]{Gill82}).  A 
straightforward balancing procedure is to project the initial data 
onto the normal modes of the system, which depends on the details of 
the numerical procedure used for forecasting (Errico~\cite{Errico91}); the 
balancing is achieved by setting the gravity wave modes to zero.  Such 
a procedure proves unsuccessful because gravity waves immediately 
appear in the numerical simulations---the nonlinear aspects of 
atmospheric dynamics are large enough to render a linear procedure 
ineffective.  Various more sophisticated initialization schemes exist 
(see Daley~\cite{Daley81}, and references therein), many of which can be 
interpreted as projecting the initial data onto a ``slow manifold'' 
(Leith~\cite{Leith80}; Lorenz~\cite{Lorenz80}), $\cal M$.  On such a slow manifold the 
gravity wave variables are given as functions of the Rossby wave 
amplitudes, and there are no rapid oscillations.


Lorenz~\cite{Lorenz80,Lorenz86} and Lorenz \& Krishnamurthy~\cite{Lorenz87} (henceforth
referred to as L80, L86 and LK87, respectively) have proposed a
series of low-dimensional models of the atmosphere and find several
difficulties with the practical computation of a slow manifold: that
various approximation schemes for $\cal M$ are divergent; that fast
gravity wave oscillations apparently must occur in some regions of
$\cal M$; and that resonances between gravity waves and Rossby waves
induce singularities in the manifold.  Such complex dynamics of such
simple models is further explored by Camassa~\cite{Camassa95} who proves, for
example, the existence of chaotic dynamics.  We concentrate here on one
of these models (L86), in which, according to the linearized governing
equations, neither Rossby waves (which we denote by $\vec x$) nor
gravity waves ($\vec y$) are subject to damping.  A three-mode
Rossby-wave complex is coupled to the two gravity-wave modes through
nonlinear interactions.  The Rossby waves have a much longer period
than the gravity waves, and in fact the period of the Rossby waves in
the model becomes infinite as their amplitude tends to zero.

We emphasize that we analyze only the low-dimensional model of L86.  
The applications of the ideas presented herein to realistic 
atmospheric equations will involve much more complicating detail.  
Nonetheless, we anticipate no great difficulty in extending the 
analysis to more physical dynamics.  The procedures utilized in this 
paper to construct slow manifolds and to appropriately initialize data 
have already been generalized successfully from low-dimensional toys 
to high-dimensional physical problems (Mercer \& Roberts~\cite{Mercer90,Mercer94a}; Roberts~\cite{Roberts93}; and Watt \emph{et 
al}~\cite{Watt95}).  The corresponding analysis of actual atmospheric 
models is left to further work.

The absence of damping in the model of L86 is a source of difficulties 
that do not arise when the waves are slightly damped, as LK87 
considered.  One might expect therefore that since atmospheric waves 
clearly are damped, the less troublesome case would be the most 
profitable to explore.  Further, the limit as the damping 
$a\rightarrow0$ is singular, and theoretically the nature of $\cal M$ 
is different in the two cases $a=0$ and $a\neq0$.  It might seem 
perverse, therefore, to tackle the undamped case.  However, a closer 
examination reveals that the mathematical difficulties do not 
disappear when damping is introduced; they are merely obscured.  For 
if $a$ is assumed to be small then inverse powers of $a$ occur in 
certain calculations, and these are a precursor of the problems that 
arise when $a=0$.  By examining the undamped case (as does Camassa~\cite{Camassa95}) we confront head-on the fundamental difficulties 
which result from resonances between waves of different timescales.

We view a rational balancing procedure as having two stages.  First, 
the functional form of the slow manifold must be determined---usually 
in the form $\vec y=\vec h(\vec x)$.  Once this is accomplished a 
closed set of low-dimensional evolution equations for the slow 
variables follows.  This reduced set could be integrated forwards in 
time to make forecasts, although in practice the full system is 
integrated from the balanced initial data.  Secondly, the appropriate 
initial values for the slow variables on $\cal M$ must be calculated 
from the full set of initial values.  Once the slow variables $\vec x$ 
are known, the appropriate fast variables are then given by $\vec 
y=\vec h(\vec x)$.  The first stage has received most attention (Baer 
\& Tribbia~\cite{Baer77}; Leith~\cite{Leith80}), while the second 
seems relatively ignored in the meteorological literature, although 
the appropriateness of some different projections, according to the 
quality of one's data measurements, has been discussed by Daley~\cite{Daley80,Daley81}.  In Section~\ref{s2} we propose a different criterion 
than those previously considered for the selection of an initial point 
on the slow manifold: that the subsequent evolution corresponds as 
closely as possible to that of the full initial-value problem, but 
with gravity waves absent.  (We shall make the notion of 
``corresponding'' evolution more precise later.)  A similar choice is 
known in physics, where fast variables are eliminated (Haake \& 
Lewenstein~\cite{Haake83}; van Kampen~\cite{vanKampen85}), and seems 
to have an obvious desirability in numerical weather forecasting: the 
adjustment of the initial data should not alter the forecast.  Many of 
the approximation schemes to compute the slow manifold implicitly 
assume that only the fast variables require adjustment, although Daley 
illustrates how this is not the case for his ``optimal'' projection 
schemes.  For our proposed projection criterion also, it turns out 
that both fast and slow variables must be altered.

In Section~\ref{s2} we briefly describe the formal computation of a slow
manifold, $\cal M$, for L86, together with the choice of appropriate 
initial conditions on $\cal M$.  In Section~\ref{s3} we discuss the divergence
of the power series for $\cal M$, and consider Lorenz' computation of
an alternative slow manifold, the ``slowest invariant manifold''
(\SIM). We find that the {\SIM} and $\cal M$ differ by terms smaller
than any power of $\vec x$ as $\vec x\rightarrow\vec0$, so that the {\SIM}
has a divergent power series.  A normal form transformation is
presented in Section~\ref{s4}, where the model L86 is written as the slow
evolution of five new variables; the five variables include the
amplitude and phase of the gravity waves.  The normal form for L86
has dynamical behavior that differs from that of L86, albeit by an
exponentially small amount. Although these differences are quantitatively
asymptotically small, we find that some solutions are qualitatively affected
by them.  Furthermore, the normal form shows that the evolution of the
slow modes {\em cannot} be entirely decoupled from the fast modes.
Thus there must be inevitable discrepancies in the long-term evolution
of the slow waves, of the order of the square of the fast waves,
between the balanced and unbalanced simulations.

Finally, in Section~\ref{s5} we review the notions of a fuzzy 
manifold, and the implications of the present work for the 
initialization of numerical weather forecasting schemes.

\section{Quasi-geostrophy as a subcentre manifold}
\label{s2}

\subsection{Computation of a slow manifold for L86}
\label{s2a}

In the model of L86, the slow ``Rossby wave'' variables $\vec x=(U,V,W)$ and
the fast ``gravity wave'' variables $\vec y=(X,Z)$ evolve according to
\begin{eqnarray}
\dot{U} & = & -VW+bVZ\nonumber\\
\dot{V} & = & UW-bUZ\nonumber\\
\dot{W} & = & -UV\label{L86}\\
\dot{X} & = & -Z\nonumber\\
\dot{Z} & = & X+bUV\nonumber,
\end{eqnarray}
where the over-dot denotes differentiation with respect to time, $b$ is a
coupling parameter, and the superscript $T$ denotes the transpose of a
row vector.  Observe that when these equations are linearized about
the zero equilibrium the Rossby wave variables remain constant, while
the gravity wave variables oscillate sinusoidally with a period that
has been normalized to $2\pi$.
 
When $b=0$ the Rossby waves and the gravity waves are uncoupled:  the
gravity waves oscillate sinusoidally with period $2\pi$; while the
Rossby waves, given by Jacobian elliptic functions, oscillate with a
period inversely proportional to their initial amplitude.  For
small-amplitude motions the Rossby waves have a much longer period than
the gravity waves---this remains superficially true when coupling is restored
($b\neq0$).

General solutions of~(\ref{L86}) have  gravity-wave components; we should 
like to adjust a given initial condition for~(\ref{L86}), 
called balancing, so that gravity waves do not develop, 
while maintaining essentially the same
evolution  of the physically significant Rossby waves.  
That is, we propose the principle that
a long-term forecast should be unaffected by the initialization.  

A first attempt to balance initial data might be projection onto the 
``geostrophic manifold'', given by $\vec y=\vec0$, that is, $X=Z=0$.  
This manifold is certainly free of gravity waves, but is not invariant 
under~(\ref{L86}); if we set $X=Z=0$ initially, they do not remain 
zero.  A more sophisticated approach to balancing is to seek an 
invariant ``quasi-geostrophic manifold'', or slow manifold, $\cal M$, 
on which solutions of~(\ref{L86}) evolve slowly.  Instead of seeking 
$\vec y=\vec0$, we allow the fast variables to be functions of the 
slow variables, $\vec y=\vec h(\vec x)$ (Leith~\cite{Leith80}).  The 
slow manifold is a so-called subcentre manifold (Kelley~\cite{Kelley67}).  Little is known about the conditions for the 
existence of subcentre manifolds; indeed resonances plague attempts to 
construct such manifolds (Sijbrand~\cite{Sijbrand85}).  In this 
section we proceed with a formal construction of $\cal M$, 
leaving the issues of its existence and uniqueness until later 
sections.

The details of the calculation of $\cal M$ are given by
Roberts~\cite{Roberts89b}.  By substituting the ansatz $\vec y=\vec
h(\vec x)$ into~(\ref{L86}), and applying the chain rule $\dot{\vec
y}=\partial \vec h/\partial \vec x\, \dot{\vec x}$, we obtain a
quasi-linear partial differential equation for $\vec h(\vec x)$
(Carr~\cite{Carr81}).  Since this PDE cannot be solved exactly we
expand $\vec h(\vec x)$ as a power series in the slow variables $\vec
x$, and find
\begin{equation}
X= -bUV+\Ord{|\vec x|^4}, \quad Z= b(U^2-V^2)W+\Ord{|\vec x|^5},
\label{XZ}
\end{equation}
as $|\vec x|\rightarrow0$.  Therefore on $\cal M$ the Rossby waves
evolve according to the following (slow) amplitude equations obtained
by substituting~(\ref{XZ}) into~(\ref{L86})
\begin{eqnarray}
\dot{U} & = & -VW\left[1-b(U^2-V^2)\right]+\Ord{|\vec x|^6}\nonumber\\
\dot{V} & = &  UW\left[1-b(U^2-V^2)\right]+\Ord{|\vec x|^6} \label{xdot}\\
\dot{W} & = & -UV\nonumber.
\end{eqnarray}
The procedure we have used to compute $\vec h(\vec x)$, that is, to 
find $X(U,V,W)$ and $Z(U,V,W)$, is mechanical, and has also been 
described for the linearly damped model L80 by Vautard and Legras~\cite{Vautard86}.  It is equivalent to the balancing scheme of Baer \& 
Tribbia~\cite{Baer77}.

\subsection{Initialization: projection of initial conditions}

We must supplement our computation of the functional form of
the slow manifold $\vec y=\vec h(\vec x)$ by deriving
appropriate initial conditions for $\vec x$ on $\cal M$.  By
``appropriate'' we mean that the subsequent evolution on $\cal M$
faithfully follows the behavior of the full system~(\ref{L86}) from
the original initial conditions.  In general it is not sufficient to
adjust only the amplitudes of the gravity waves, that is, to map the
initial point $(\vec x^*,\vec y^*)$ of the full system to the point
$(\vec x^*,\vec h(\vec x^*))$ on $\cal M$: if the full system and the
slow-manifold model are to have the same future behavior from their
respective initial conditions, the initial values of the Rossby wave variables also
must be adjusted.  The discrepancy between the initial condition for
the slow variables in the full initial-value problem and that on the
slow manifold is known as ``initial slip'' (Grad~\cite{Grad63}), by analogy
with the slip allowed for an inviscid fluid at a boundary.  We now
proceed to calculate this ``initial slip''.

The appropriate choice of an initial point on $\cal M$ has a
straightforward derivation (Roberts~\cite{Roberts89b}) when $\cal M$ attracts
neighboring solutions exponentially.  In that case, the choice may be
made so that the solution from the original initial condition, and that
from the adjusted initial condition approach one another exponentially
quickly as $t\rightarrow\infty$.  However, in the present case the slow
manifold does not attract neighboring trajectories, but instead acts
as a centre for their gravity-wave oscillations---a solution initially
off $\cal M$ oscillates about $\cal M$ perpetually.  We therefore aim
to choose the initial point on $\cal M$ so that its subsequent
evolution maintains its relationship with the full uninitialized
solution for all time.

The procedure we describe below is correct to leading order in the
distance, $r$, of the initial point $(\vec x^*,\vec y^*)$ from $\cal
M$.  In Section~\ref{s4} we shall show that in general it is not possible to
improve the choice of initialized point.  For example, we expected to
be able to incorporate corrections of $\Ord{r^2}$; however, we find that
this cannot be done and so solutions on and off $\cal M$ necessarily
drift apart after a time of $\Ord{r^{-2}}$.

In order to explain our procedure we begin by briefly considering
the simple projection
\begin{equation}
(\vec x^*,\vec y^*)\mapsto (\vec x^*,\vec h(\vec x^*)),
\label{BTp}
\end{equation}
used implicitly by Baer \& Tribbia~\cite{Baer77}.  The adjustment of initial
data, by the displacement $(\vec 0,\vec h(\vec x^*)-\vec y^*)$, lies in
the ``gravity-wave'' space spanned by the vectors $(0,0,0,1,0)$ and
$(0,0,0,0,1)$: only the gravity-wave variables are altered. We
call the solution obtained by projecting initial data in this way BT. However,
for the projection we propose here, the two vectors that span the appropriate
direction for the projection of initial conditions depend on $\vec x^*$
(Roberts~\cite{Roberts89b}), and in general the Rossby-wave variables are also changed
by the initialization process.
The simple projection~(\ref{BTp}) is appropriate only in the special case
when $\vec x^*=\vec 0$, where we may observe that $\vec x(t)=\vec 0$
for all $t>0$, according to both the original system~(\ref{L86}) and
the slow model~(\ref{xdot}).


A more informed projection scheme must be based upon the evolution near
the slow manifold (Roberts~\cite{Roberts89b}).  If $\vec U(t)$ is a
solution of~(\ref{L86}) on $\cal M$, and if
$\vec\epsilon(t)=(\epsilon_1 ,\epsilon_2 ,\epsilon_3 ,\epsilon_4
,\epsilon_5)$ is an infinitesimal displacement from $\vec U$ then
$\vec\epsilon$ evolves, according to~(\ref{L86}), as
\begin{equation}
\dot{\vec \epsilon} = \left({\cal L}+{\cal N}_1\right)\vec \epsilon,
\label{epsdot}
\end{equation}
where ${\cal L}\vec\epsilon= (0,0,0,-\epsilon_5,\epsilon_4)$, and
${\cal N}_1$ is the Fr\'echet derivative of the vector of nonlinear
terms ${\cal N}(\vec x)=(-VW+bVZ,UW-bUZ,-UV,0,bUV)$ evaluated on $\cal
M$ (see Roberts~\cite{Roberts89b}).  For any slow solution on $\cal M$,
we then seek those neighboring solutions that evolve {\em with} the
solution on $\cal M$ plus a fast oscillation.  It is just these
neighboring solutions which should be projected onto the particular
slow solution on $\cal M$ in order to maintain the long-term forecast.
\OMIT{ In particular, if $\vec \epsilon_1$ and $\vec \epsilon_2$ are
two linearly independent infinitesimal displacements in the
``projection space'', that is, in the direction appropriate for
projection of initial conditions, then both satisfy~(\ref{epsdot}).
For a complete specification of $\vec\epsilon_1$ and $\vec\epsilon_2$
we need a side condition to supplement~(\ref{epsdot}), which can be
determined by recalling that when $\vec x=\vec 0$ the projection space
is spanned by $\vec\epsilon_1(\vec 0)=(0,0,0,1,0)$ and
$\vec\epsilon_2(\vec 0)=(0,0,0,0,1)$.  In order to
solve~(\ref{epsdot}), we first note that the evolving displacements may
be written as functions of position $\vec x$ on $\cal M$, and so by the
chain rule the time derivative in~(\ref{epsdot}) becomes
\[
\dot{\vec \epsilon}_j=
\sum_{k=1}^3 \D{\vec\epsilon_j}{x_k} \dot{x}_k.
\]
}
We wish to calculate only the projection space spanned by $\vec\epsilon_1$ and
$\vec\epsilon_2$, and not their magnitudes nor their individual directions.
\OMIT{
so we decompose the vectors into the form
\begin{equation}
\left[ \vec \epsilon_1,\vec \epsilon_2\right]=
\left[ \begin{array}{c}
       P\\
       I
       \end{array}\right]
       Q\equiv
\left[ \begin{array}{cc}
       p_{11} & p_{12}\\
       p_{21} & p_{22}\\
       p_{31} & p_{32}\\
       1      & 0     \\
       0      & 1     
       \end{array}\right]
\left[ \begin{array}{cc}
       q_{11} & q_{12}\\
       q_{21} & q_{22}
       \end{array}\right],
\label{epsPQ}
\end{equation}
where $Q(\vec x)=[q_{ij}(\vec x)]$ is an invertible matrix.  The
columns of the matrix $\left[ P^T \;\; I\right]^T$ span the projection
space.  When $\vec x=\vec0$, the elements of the matrix $P(\vec
0)=[p_{ij}(\vec 0)]$ are zero, and $Q(\vec 0)$ is the identity matrix.
After substituting~(\ref{epsPQ}) into~(\ref{epsdot}) we find that we
can eliminate $Q$ to yield an equation for the matrix $P$,
\begin{equation}
\sum_{k=1}^3 \D{P}{x_k} \dot{x}_k =
\left[ \begin{array}{cc} I & -P^T \end{array}\right]
\left({\cal L} + {\cal N}_1(\vec x)\right)
\left[\begin{array}{c} P\\I
       \end{array}\right].
\label{Peqn}
\end{equation}
This quasi-linear partial differential equation determines the
projection space.  To solve~(\ref{Peqn}) Roberts~\cite{Roberts89b} developed a
power-series expansion for the matrix $P$. 
}
The argument is elaborated in detail in Section~4 of the paper
by Roberts~\cite{Roberts89b}.
The result to low-order is
that the projection onto $\cal M$ should be made along the position
dependent planes spanned by vectors $\vec e_1$, $\vec e_2$, where
\begin{equation}
\vec e_1=\left[\begin{array}{c}-bV\\bU\\0\\1\\0\end{array}\right]
+\Ord{|\vec x|^3},
\quad
\vec e_2=\left[\begin{array}{c}0\\0\\-b(U^2-V^2)\\0\\1\end{array}\right]
+\Ord{|\vec x|^3}.
\label{vecs}
\end{equation}
We shall call the solution obtained after projecting initial data along these
vectors R.

\begin{figure}
	\centerline{\includegraphics{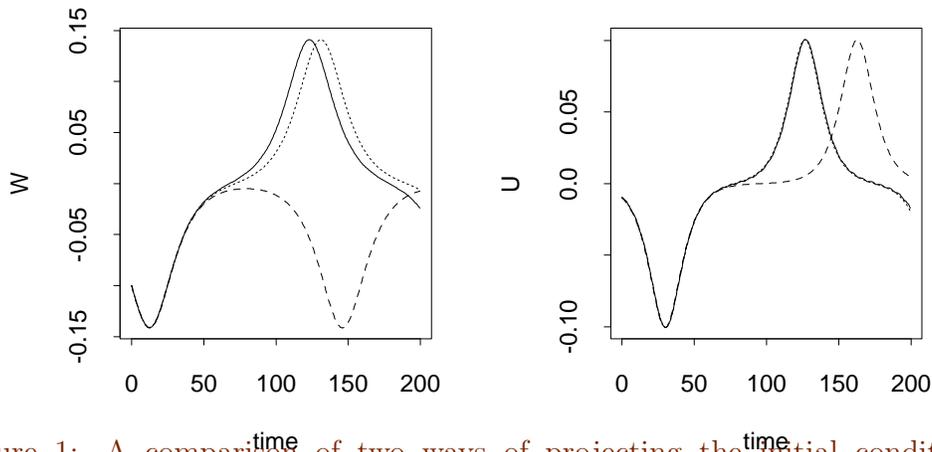}}
\caption{A comparison of two ways of projecting the initial condition
$(\protect\vec x^*,\protect\vec y^*)$ onto $\cal M$: BT and R.
Each graph shows three
solutions of (\protect\ref{L86}): the solid line from the initial condition
$(\protect\vec x^*,\protect\vec y^*)$; the dotted line from projection R;
the dashed line
from projection BT. Initial conditions are: (i)
$(\protect\vec x^*,\protect\vec y^*)\equiv(U,V,W,X,Z)=(0.1,0.1,-0.1,0.01,0.01)$;
(ii) $(\protect\vec x^*,\protect\vec y^*)=(-0.01,0.1,0.01,0.01,0)$. In graph
(ii) the solid line and the dotted line are almost indistinguishable,
except for the small-amplitude gravity-wave wiggles present in the
solid line, but not in the dashed. Projection R is clearly better than BT.
Both initialization schemes use just the first few terms in the relevant
power-series expansions, as given by (\protect\ref{XZ}) and
(\protect\ref{vecs}).}
\label{ic}
\end{figure}

In Figure~\ref{ic} we show some typical results for solutions BT and R, and
compare them with solutions of~(\ref{L86}) from uninitialized initial data.
We have chosen different initial conditions for the two graphs; these
are representative of a large number of numerical runs. 
In both cases, R  is clearly closer to the uninitialized solution than is BT.
In the first, BT and R are both phase-shifted with respect to the
original solution, but R less so than BT. Further, BT goes the
``wrong way'' round a fixed point of~(\ref{L86}) at time $t\approx50$,
and as a result the slow variable $W$ takes the wrong sign.
In the second case, R and the 
original solution are indistinguishable on the figure (although they slowly
diverge at later times); at times $t\approx75,150$ the original solution has
gravity-wave wiggles that the projected solutions do not have (because
they both lie close to the slow manifold). Here, R is strikingly
better than BT.

\subsection{Incorporation of forcing}

The centre manifold formalism allows one  to compute the effects of a
forcing of the full system~(\ref{L86}) on the dynamics of the slow
manifold (or, for that matter, the damped, forced system of LK87).
This computation is described to leading order in the amplitude of the
forcing by Cox \& Roberts~\cite{Cox91}, and shows that the forcing not only
changes the evolution of the slow ``Rossby'' wave variables, but also
changes the shape and location of a slow manifold $\cal M$.  For
example, if we add a constant forcing $\vec F=(F_U, F_V, F_W, F_X,
F_Z)$ to the right-hand side of~(\ref{L86}) then the slow manifold becomes,
to leading order in $|\vec F|$,
\begin{eqnarray}
X&=&-bUV-F_Z+(U^2-V^2)F_W-2W(VF_V-UF_U)
\nonumber\\&&{}
+\Ord{|\vec x|^4,|\vec F||\vec x|^3}
\label{n1}\\
Z&=&bW(U^2-V^2)+F_X+b(UF_V+VF_U)-b^2F_X(U^2-V^2)\nonumber\\
&&{}+\Ord{|\vec x|^5,|\vec F| |\vec x|^3}.  \label{n2}
\end{eqnarray}
We have previously shown (Cox \& Roberts~\cite{Cox91}) that when forcing is
present, it is important to project initial conditions onto a slow
manifold given by~(\ref{n1}--\ref{n2}), rather than the unperturbed
slow manifold~(\ref{XZ}), in order to eliminate gravity waves.

\section{Non-uniqueness of a slow manifold}
\label{s3}

\subsection{Divergence of series}

Lorenz~\cite{Lorenz86} demonstrates that several approximation schemes 
aimed at constructing a slow manifold yield divergent power series.  
Two of these schemes (Baer \& Tribbia~\cite{Baer77}; Vautard \& 
Legras~\cite{Vautard86}) are equivalent to the construction of a 
subcentre manifold (Roberts~\cite{Roberts89b}) that we have just 
described.  Not surprisingly then, the expansions~(\ref{XZ}) are found 
to be divergent.

Such divergence is not necessarily a computational disaster.  
Techniques such as Pad\'e summation or the Shanks transform not only 
improve the rate of convergence of a series, they may also produce a 
converged ``sum'' of a divergent series---the Stieltjes series is a 
traditional example (see Section~8.3 in Bender \& Orszag~\cite{Bender81}).  The use of such convergence-acceleration techniques 
in fluid dynamics has been reviewed by Van~Dyke~\cite{Vandyke84}.  Here the 
series expansion~(\ref{XZ}) for $\cal M$ is in terms of the three slow 
variables $U$, $V$ and $W$.  However, the acceleration of convergence 
of such multi-variate expansions is less well understood than that for 
expansions in a single variable.  Perhaps the best current technique 
is to use the multi-variate Pad\'e transform proposed by Cuyt~\cite{Cuyt84}, 
but our preliminary investigations are so far inconclusive.  At the 
very least, low-order computations of the shape of a slow manifold can 
be summed to within an exponentially small error using an ``optimal 
truncation'' of the series (Bender \& Orszag~\cite[pp.94--100,122--123]{Bender81}).

\subsection{Construction of the {SIM}}

Lorenz (L86) describes an alternative construction of a slow manifold,
called the ``slowest invariant manifold'' ({\SIM}), for which convergence
is not a problem.  The calculation proceeds by generating a family of
periodic solutions to~(\ref{L86}), using a numerical shooting method to
determine the correct initial value of one of the variables in order to
ensure periodicity.  This family forms an invariant manifold, the
{\SIM}, and the central issue is whether such an object is free of
significant gravity-wave activity.  The construction is heavily
dependent on the symmetries of~(\ref{L86}).

Lorenz' method of generating the periodic orbits is to consider
solutions for which both $V^*\equiv V(0)$ and $X^*\equiv X(0)$
are initially zero.  He then takes initial values $W^*>U^*>0$
and treats the remaining initial condition, $Z^*$, as a parameter
to be chosen so that at the first zero-crossing of $U$, when $t=T$
say, $X$ vanishes too. A consequence of the simultaneous vanishing
of $U(T)$ and $X(T)$, together with the symmetries of~(\ref{L86}) (see L86),
is that the constructed solution is periodic, with period $4T$.
For a fixed ratio $m^{1/2}\equiv U^*/W^*$, Lorenz computes a
one-parameter family of periodic orbits, as $W^*$ is varied. 
The curve $Z=Z^*(W^*)$ is then a section through the
{\SIM}, and the entire {\SIM} may in principle be computed numerically
by varying the ratio $U^*/W^*$.  Lorenz finds that the {\SIM} contains
an infinite number of singularities, which are increasingly closely
packed as $W^*\rightarrow0$.  He concludes that ``there is no
unequivocally slow manifold'' because his candidate slow manifold, the
{\SIM}, contains regions of high gravity-wave activity close to the
singularities.

The central result of this section will be that Lorenz' {\SIM} and the
slow manifold $\cal M$ computed in the previous section are
exponentially close, a concept we elaborate upon below.  Furthermore,
a slow manifold is not unique: indeed, there are infinitely many
pretenders to that title. Such nonuniqueness arises frequently in
the construction of low-dimensional models of dynamical systems
(Roberts~\cite{Roberts89}), and each has the same (divergent) power-series expansion.

Lorenz shows how the singularities in the {\SIM} arise from resonances
between the slow Rossby waves and the fast gravity waves, by considering
the uncoupled system~(\ref{L86}) with $b=0$.  Then the section through
the {\SIM} in the $(W^*,Z^*)$-plane consists of the horizontal line
$Z^*=0$, on which there is no gravity-wave activity, together with the
vertical lines $W^*=K/(k\pi)$, for integers $k$, on which the period of
the Rossby waves is a multiple of the gravity-wave period. 
The singularities of the {\SIM} when $b\neq0$ come from a
perturbation of the structure of the uncoupled system.  Lorenz noted that the
singular curve $Z=Z^*(W^*)$ looked like $Z=a \exp(1/W^*)\cot(K/W^*)$.
By computing the analytic form of $Z^*$, up to terms of $\Ord{b}$, we now
show that this expression for $Z^*$ is qualitatively correct, and that
the singularities are indeed exponentially weak as $W^*\rightarrow0$.
To do so we expand the variables as power series in the coupling
parameter $b$, so that
\[
U\sim\sum_{j=0}^{\infty} b^jU_j,
\]
and so on.  (LK87 make a similar expansion for the variables, although
with a different purpose.) We take initial conditions
\[
U(0)=U^*,\quad V(0)=0,\quad W(0)=W^*,\quad X(0)=0,\quad
Z(0)=Z^*\sim\sum_{j=0}^\infty b^jZ^*_j,
\]
where $U^*$ and $W^*$ are fixed independently of $b$, and we let the
time of the first zero-crossing of $U(t)$ be at $t=T=T_0+bT_1+\cdots$.
Recall that in order to find periodic solutions of~(\ref{L86}), our aim
is to find $T$ and $Z^*$ so that
\begin{equation}
U(T)=X(T)=0.
\label{UTXT0}
\end{equation}
At leading order in $b$ the variables satisfy
\begin{eqnarray}
\dot{U}_0 & = & -V_0W_0 \nonumber\\
\dot{V}_0 & = &  U_0W_0\nonumber\\
\dot{W}_0 & = & -U_0V_0 \label{b0}\\
\dot{X}_0 & = & -Z_0\nonumber\\
\dot{Z}_0 & = &  X_0\nonumber.
\end{eqnarray}
The solution (L86) is
\[
U_0=U^* \CN (W^*t),\quad
V_0=U^* \SN (W^*t),\quad
W_0=W^* \DN (W^*t),
\]
\[
X_0=-Z_0^* \sin t,\quad
Z_0=Z_0^* \cos t,
\]
where \CN, \SN, {\DN} are Jacobian elliptic functions (Abramowitz and
Stegun, Chapter 16) with parameter $m=\left(U^*/W^*\right)^2$.

\begin{figure}
\centerline{\includegraphics{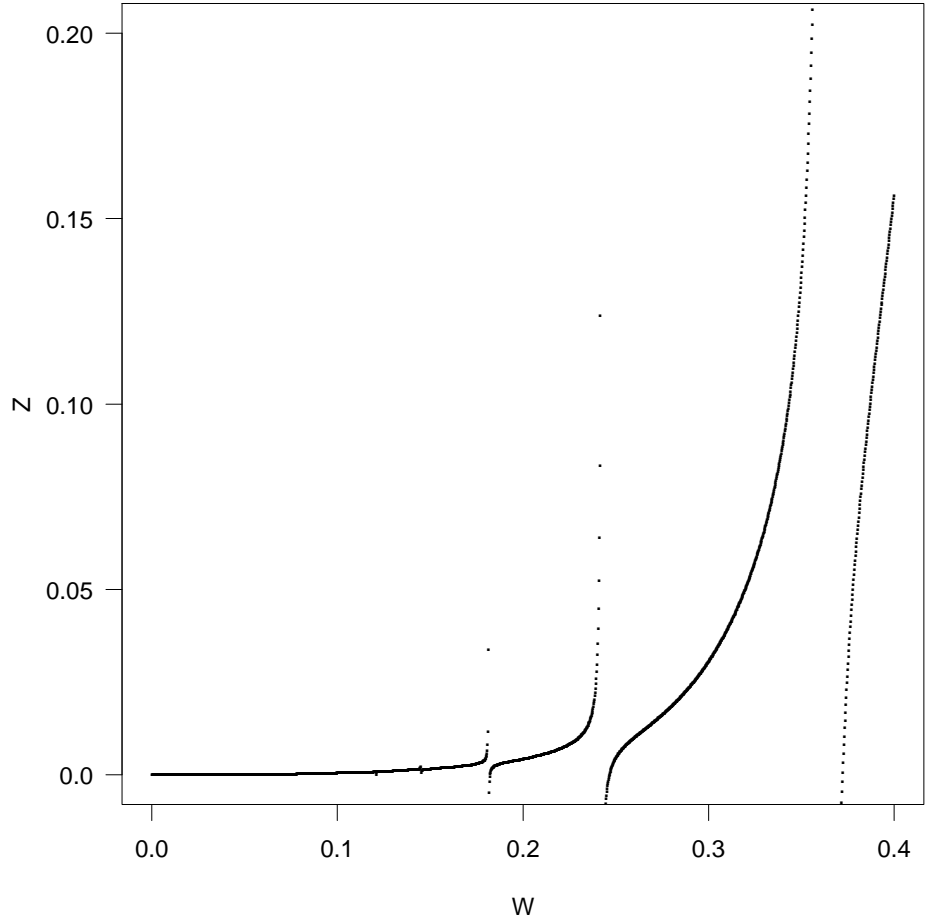}}
\caption{A section through the \SIM\ given by the analytic
expression~(\protect\ref{Z1*}).}
\label{figsim}
\end{figure}

Expanding~(\ref{UTXT0}) in powers of $b$, we find at leading order that
$T_0$ satisfies $U_0(T_0)=0$.  The first zero-crossing of $U_0(t)$
occurs when $t=T_0=K/W^*$, where
\begin{equation}
K=
K(m)=\int_{0}^{\pi/2}\frac{d\theta}{\left(1-m\sin^2\theta\right)^{1/2}}
\label{Km}
\end{equation}
is the complete elliptic integral of the first kind.
Thus $X_0(T_0)=-Z_0^*\sin (K/W^*)$, which vanishes if either $Z_0^*=0$
or if $W^*=K/(k\pi)$, for some integer $k$.  This is Lorenz' result for
$b=0$.  The line $Z_0^*=0$ is the flat approximation to the {\SIM}, and
the vertical lines $W^*=K/(k\pi)$ indicate resonances between the
Rossby waves and the gravity waves.  We take $Z^*_0=0$, so that
$X_0=Z_0=0$, and now consider the terms of $\Ord{b}$ in the expansion of
the {\SIM}.  These satisfy
\begin{eqnarray}
\dot{U}_1 & = & -V_0W_1-V_1W_0\nonumber\\
\dot{V}_1 & = &  U_0W_1+U_1W_0\nonumber\\
\dot{W}_1 & = & -U_0V_1-U_1V_0\label{b1}\\
\dot{X}_1 & = & -Z_1\nonumber\\
\dot{Z}_1 & = &  X_1+U_0V_0\nonumber,
\end{eqnarray}
subject to
\[
U_1(0)=V_1(0)=W_1(0)=X_1(0)=0, \quad Z_1(0)=Z^*_1.
\]
We see immediately that $U_1=V_1=W_1=0$ (and indeed, 
$U_{2n+1}=V_{2n+1}=W_{2n+1}=X_{2n}=Z_{2n}=0$ for all $n$).  However,
the equations for the gravity waves yield
\begin{equation}
X_1(t)=-Z_1^* \sin t-{U^*}^{2}\int_0^t \sin(t-\tau)\, \CN (W^*\tau)\,
\SN (W^*\tau)\, d\tau \ ,
\label{X1}
\end{equation}
and $Z_1(t)=-\dot{X}_1(t)$ from~(\ref{b1}).  Now we examine~(\ref{UTXT0}) at $\Ord{b}$, which yields
\[
T_1 \dot{U}_0(T_0)=0, \quad X_1(T_0)=0.
\]
It follows that $T_1=0$, and upon substituting~(\ref{X1}) into the
second condition, we must choose $Z_1^*$ so that\footnote{This result 
may also be derived by considering the terms of $\Ord{b}$ as a forcing
of the unperturbed system, as in Cox \& Roberts~\cite{Cox91}.}
\begin{equation}
Z_1^* \sin T_0 = -{U^*}^{2}\int_{0}^{T_0} \sin(T_0-\tau)\,
\CN (W^* \tau)\, \SN (W^* \tau)\, d\tau.
\label{goApp}
\end{equation}
By contour integration (see the appendix for details) we find
\begin{equation}
Z_1^* = \frac{\pi}{2} \sech\left(\frac{K'}{W^*}\right)
\cot\left(\frac{K}{W^*}\right)+
W^*F(W^*),
\label{Z1*}
\end{equation}
as shown in Figure~\ref{figsim},
where $K'=K(1-m)$ and where $F(W^*)$ has the asymptotic series
\begin{eqnarray}
F(W^*) & \sim &  \sum_{n=1}^{\infty} \DC^{(2n)}(0|1-m){W^*}^{2n}\nonumber\\
 & \sim & m{W^*}^{2} + m(m+4){W^*}^{4}+m(m^2+44m+16){W^*}^{6}\nonumber\\
 & & {}+ m(m^3+408m^2+912m+64){W^*}^{8}+\Ord{{W^*}^{10}}.
\label{F}
\end{eqnarray}
Here $\DC^{(2n)}$ is the $2n$-th derivative of the Jacobian elliptic
function $\DC$.

Observe that through the factor of $\cot (K/W^*)$ the behavior of
$Z^*_1$ is singular for $W^*$ near $K/(k\pi)$, but there is an
exponentially small factor, $\sech(K'/W^*)$, multiplying this singular
term; as $W^*\rightarrow0$ the singularity rapidly becomes weaker.  Our
approximation of the {\SIM} to $\Ord{b}$, given by~(\ref{Z1*}), agrees
very well with the {\SIM} computed numerically by Lorenz for $b=0.5$.
He finds $Z^*\approx 0.4 {W^*}^{3}$ for small $W^*$ for an initial
condition corresponding to $m=0.81$, while our equivalent result is
$Z^*\approx mb{W^*}^{3}= 0.405 {W^*}^{3}$.

\subsection{Divergent power series and periodic solutions on the {SIM}
and on $\cal M$}

We now make a few observations about the expression~(\ref{Z1*}) for the
section through the {\SIM}.  Undoubtedly corrections will be made to
$Z^*$ at orders $b^3$, $b^5$, \ldots, but for our present purposes we
assume that the essential features of the curve $Z^*(W^*)$ are captured
by~(\ref{Z1*}) alone.

Firstly, we recall the result that, in general, centre manifolds (and 
subcentre manifolds, as are at issue here) are not unique.  A given 
system may have an infinite number of centre manifolds, each differing 
from any other by exponentially small terms (see Carr~\cite{Carr81}; 
and Roberts~\cite[\S2.1]{Roberts85b}).  Each has an identical 
power-series representation.  Therefore the existence of more than one 
slow manifold ($\cal M$, and the {\SIM}, and infinitely many others) 
for the atmospheric model L86 should come as no surprise.  The notable 
feature of the exponentially small difference between $\cal M$ and the 
{\SIM} is its multiplication by a term with infinitely many algebraic 
singularities, $\cot(K/W^*)$.  This frustrating feature results from 
resonances between the fast and the slow waves that arise in 
attempting to construct a subcentre manifold, rather than a centre 
manifold (see Sijbrand~\cite{Sijbrand85}).  Frequently, when one 
computes the power series for a centre manifold, it is divergent, even 
though a centre manifold contains no singularities similar to those of 
a subcentre manifold.  We wish, then, to emphasize that the existence 
of singularities in the {\SIM} and the divergent power series for any 
slow manifold are separate issues.

Secondly,~(\ref{Z1*}) indicates that the {\SIM} and the slow manifold
$\cal M$ differ by terms that are exponentially small as
$W^*\rightarrow0$.  To see this we use the fact that we have chosen $V^*=0$
in our search for periodic solutions of~(\ref{L86}), and so according to
the slow-manifold model, $Z^*=Z(U^*,0,W^*)$.
Using Roberts'~\cite{Roberts89b} expansion of $Z(U,V,W)$, (\ref{XZ}), we find
\[
Z^*=bW^*{U^*}^{2}+bW^*{U^*}^{2}(4{W^*}^{2}+{U^*}^{2}) + \cdots + \Ord{b^2}.
\]
This agrees with the result~(\ref{Z1*}) when the exponentially small term is
ignored, since ${U^*}^{2}=m{W^*}^{2}$.  Indeed, further investigation reveals
agreement between $W^*F(W^*)$ and $Z(U^*,0,W^*)$ at all orders in $|(U^*,W^*)|$.

Thirdly, let us examine more closely the result of using the
slow-manifold model~(\ref{XZ}--\ref{xdot}) to approximate the \SIM\
of~(\ref{L86}).  We begin by noting that the construction of $\cal M$
yields expressions for $X(U,V,W)$ and $Z(U,V,W)$ of the form
\[
X=UV\tilde{f}(U,V,W),\quad
Z=bW\tilde{g}(U,V,W).
\]
It follows that a zero-crossing of either $U$ or $V$ necessarily implies a
zero-crossing of $X$.  Further, the evolution of the slow variables $U$,
$V$, $W$ on $\cal M$ is governed by equations of the form
\begin{eqnarray}
\dot{U}&=&-VW(1-b^2\tilde{g})\nonumber\\
\dot{V}&=&UW(1-b^2\tilde{g})\label{3d}\\
\dot{W}&=&-UV\nonumber,
\end{eqnarray}
so that $U^2+V^2$ is invariant for this model, just as it is for the
original system (L86).  Consequently, the dynamics of~(\ref{3d}) are
two-dimensional and so its solutions are in general periodic. Since, by
definition, the \SIM\ is composed of the periodic solutions of~(\ref{L86}),
it might therefore appear that $\cal M$ {\em is} the \SIM\ of~(\ref{L86}).
However, we know from our preceding remark that $\cal M$ and the
\SIM\ of~(\ref{L86}) are {\em not} the same---they differ by exponentially
small terms. This discrepancy is resolved by concluding that a slow
manifold $\cal M$, constructed by the functional relationship $\vec
y=\vec h(\vec x)$, cannot be exactly invariant under evolution
according to~(\ref{L86})---there will always remain exponentially small
discrepancies.

The \SIM\ was proposed because it can be computed directly without
recourse to (divergent) power series.  However, the power series for
the \SIM\ is the same as for $\cal M$, and is divergent.  We emphasize
that the divergence of its power series does not mean that the \SIM\
does not exist, nor does it mean that the \SIM\ is ill-defined, or
``fuzzy''.  The utility of the divergent series is illustrated by the
excellent agreement between an approximation to $Z^*(W^*)$ based on a
small number of terms for $\cal M$, and the numerical result for the
behavior of $Z^*(W^*)$ as $W^*\rightarrow0$ on the \SIM.  For
divergent series, this agreement is good for small $W^*$ and for a
fixed number of terms.  In the next section we shall illustrate some of
the dynamical consequences of the differences between $\cal M$ and the
\SIM.  For most solutions the differences are slight, and result in
small quantitative changes. In a practical computation, we are required
to select one of the many possible slow manifolds. We consider two strong
advantages of $\cal M$ over the \SIM\ to be the possibility of computing
appropriately the projection of initial conditions onto $\cal M$,
and the explicit algebraic form $\vec y=\vec h(\vec x)$ of $\cal M$,
compared with the construction of the \SIM\ through an ensemble of numerical
solutions.


\newpage
\section{Normal form for L86}
\label{s4}

In Section~\ref{s2} we described the construction of a slow manifold,
$\cal M$, which approximates a subset of the possible solutions
of~(\ref{L86}), and on which the fast ``gravity'' wave variables are
slaved to the amplitudes of the slow ``Rossby'' waves.  More general
solutions of~(\ref{L86}), with significant independent fast wave
activity, involve oscillations centreed on $\cal M$.  In this section
we rewrite~(\ref{L86}) so that all solutions are captured, not just the
slow solutions on $\cal M$, but where the new governing equations
describe the evolution of five new slowly-varying quantities.  This is
a normal form calculation (see Guckenheimer \&
Holmes~\cite{Guckenheimer83}; Arnold~\cite{Arnold82}), which has been
sketched for the forced, damped system of LK87 by
Jacobs~\cite{Jacobs91}.  In that case the variables $U$, $V$, $W$, $X$,
$Z$ apparently may be written as {\em convergent} power series in the
new slow variables: for the model L86, where there is no damping, we
shall see that the equivalent series are divergent.  Jacobs notes that
in the new variables a slow manifold takes a particularly simple form.
We shall also see from the normal form calculation that there is a
limit on our ability to initialize data in such a way that the
long-time dynamics of the physically dominant slow ``Rossby'' waves are
unaffected by the balancing procedure.  That is, there is a limit on
the agreement between the unbalanced and the balanced systems---the two
forecasts inevitably diverge slowly.

We begin by noting that for the uncoupled system, (\ref{L86}) with
$b=0$, we can identify five slowly-varying quantities: $U$, $V$, $W$,
$R$ and $\dot\Theta$, where $X=R\cos\Theta$ and $Z=R\sin\Theta$.  (In
fact, when $b=0$ it follows that $\dot{R}=0$ and $\dot{\Theta}=1$.) Our
aim now is to find for the coupled system, (\ref{L86}) with $b\neq0$,
five equivalent slowly varying quantities.  We do this by making
successive nonlinear changes of variables: to leading order in the
calculation the slowly varying quantities will be just $U$, $V$, $W$,
$R$ and $\dot\Theta$.  We shall perform algebraic manipulations on the
system of equations, but there is no reduction in the dimension of
the system as there was in computing a slow manifold $\cal M$; we expect
the normal form to capture all solutions of~(\ref{L86}).  This point is the
essential difference between the  simplifying tools of invariant
manifold theory and normal form theory. The first aims to reduce the
dimension of a nonlinear dynamical system, while the second transforms
the system to a canonical form.

To make the normal form transformation we seek to write~(\ref{L86}) in
terms of slowly varying variables $u$, $v$, $w$, $r$ and
$\dot{\theta}$. Linearly, these will be identified with the original
variables $U$, $V$, $W$, $R$ and $\dot\Theta$, respectively.  Thus the
evolution of $u$, $v$ and $w$ will describe predominantly the dynamics
of the slow  waves, while $r$ and $\theta$ represent the amplitude
and phase of the fast  waves.  We shall expand the ``physical'' variables
$(U,V,W,X,Z)$ as power series in the new variables $\vec u=(u,v,w,x,z)$,
where the two variables $x=r\cos\theta$ and
$z=r\sin\theta$ are closely related to the original fast wave
variables $X$ and $Z$.  This power-series expansion introduces
exponentially small errors; the new evolutionary system for $\vec u$
will have solutions that differ from solutions of~(\ref{L86}) by terms
smaller than any power of $|\vec u|$ as $|\vec u|\rightarrow0$.
Further, if for a practical computation we truncate the transformation
at a finite number of terms in the power series, then larger, algebraic
errors are introduced.

Recall that $\vec x=(U,V,W)$ and $\vec y=(X,Z)$ so that~(\ref{L86}) may
be written as the following pair of equations,
\begin{eqnarray}
\dot{\vec x} & = & \vec f(\vec x,\vec y)\label{4.1a}\\
\dot{\vec y} & = & B\vec y+\vec g(\vec x),\label{4.1b}
\end{eqnarray}
where 
\[
\vec f(\vec x,\vec y)=(-VW+bVZ,UW-bUZ,-UV),\quad
\]
\[
B=\left[\begin{array}{cc} 0 & -1\\1 & 0\end{array}\right],\quad
\vec g(\vec x)=(0,bUV).
\]
We make a change of variables,
\begin{equation}
\vec x = \vec\chi + \vec F(\vec\chi,\vec\eta), \qquad
\vec y = \vec\eta + \vec G(\vec\chi,\vec\eta),
\label{4.2}
\end{equation}
where we identify $\vec\chi=(u,v,w)$ and $\vec\eta=(x,z)$.  Here $\vec
F$ and $\vec G$ are $O(|(\vec\chi,\vec\eta)|^2)$ as
$|(\vec\chi,\vec\eta)|\rightarrow0$, so that linearly $\vec
x\sim\vec\chi$ and $\vec y\sim\vec\eta$.  Substitution of~(\ref{4.2})
into~(\ref{4.1a}--\ref{4.1b}) gives the following equations for the
evolution of the new variables:
\begin{eqnarray}
\dot{\vec\chi}+\D{\vec F}{\vec\chi}\dot{\vec\chi}+
\D{\vec F}{\vec\eta}\dot{\vec\eta}
 & = & \vec f(\vec\chi+\vec F,\vec\eta+\vec G) \label{4.3a}\\
\dot{\vec\eta}+\D{\vec G}{\vec\chi}\dot{\vec\chi}+
\D{\vec G}{\vec\eta}\dot{\vec\eta}
 & = & B\vec\eta+B\vec G+\vec g(\vec\chi+\vec F).\label{4.3b}
\end{eqnarray}
We now assume that the nonlinear terms and the new variables' evolution
may be written as the power series
\begin{eqnarray}
\vec F\sim\sum_{n=2}^\infty\vec F_n,&\qquad&
\vec G\sim\sum_{n=2}^\infty\vec G_n,
\label{FG}
\\
\dot{\vec\chi} \sim \sum_{n=2}^\infty \vec R_n(\vec\chi,\vec\eta),&\qquad&
\dot{\vec\eta} \sim B\vec\eta+ \sum_{n=2}^\infty \vec S_n(\vec\chi,\vec\eta),
\label{RS}
\end{eqnarray}
where $\vec F_n$, $\vec G_n$, $\vec R_n$, $\vec S_n
=O(|(\vec\chi,\vec\eta)|^n)$.  We shall solve~(\ref{4.3a}--\ref{4.3b})
at successive orders in $|(\vec\chi,\vec\eta)|$; at each order our aim
is to choose the four quantities $\vec F_n$, $\vec G_n$, $\vec R_n$,
$\vec S_n$ as simply as possible. We shall see below what the term ``simple'' actually
means.  It turns out to be possible at each order to require that
$\dot{\vec\chi}$, $\dot{r}$ and $\dot{\theta}$ are slowly varying, that
is, that they are independent of the phase of the fast waves, $\theta$.

For use in the calculation below, we note that by the chain rule
\[
\D{a}{\theta}=\D{a}{x}\D{x}{\theta}
             +\D{a}{z}\D{z}{\theta}
             =-\D{a}{x}z
              +\D{a}{z}x
             =\D{a}{\vec\eta}B\vec\eta,
\]
for any function $a(u,v,w,x,z)$.  As an example of some of the algebraic
details, consider the equation governing the slow  wave components
$\vec R_2$ and $\vec F_2$ which is, from~(\ref{4.3a}--\ref{RS}),
\[
\vec R_2=\vec f_2 -\D{\vec F_2}{\vec\eta}B\vec\eta
        =\vec f_2 -\D{\vec F_2}{\theta},
\]
where
\[
\vec f_2=(-v(w-bz),u(w-bz),-uv).
\]
Note that $\vec f_2$ depends on $\theta$ through $z$.  We now choose
$\vec R_2$ to be independent of $\theta$ {\em and} such that secular
growth in $\vec F_2$ is avoided. This requires
\[
\vec R_2= (-vw,uw,-uv),
\]
in which case
\begin{equation}
\D{\vec F_2}{\theta}=(bvz,-buz,0).
\label{dfdtheta}
\end{equation}
Now we choose $\vec F_2$ as simply as possible, that is,
$\vec F_2=(-bvx,bux,0)$. 
We can add to this expression for $\vec F_2$ an arbitrary
function of $u$, $v$, $w$ and $r$, while still satisfying~(\ref{dfdtheta}),
but for simplicity we decide that $\theta$-averages are to vanish.
A consequence of this decision is that after averaging over the gravity
waves $u$, $v$ and $w$ reduce to $U$, $V$ and $W$ respectively.

For general $n>2$,
\begin{eqnarray*}
\vec R_n & = & 
     \vec f_n -\sum_{m=2}^{n-1}\D{\vec F_m}{\vec\chi}\vec R_{n+1-m}
     -\sum_{m=2}^{n-1}\D{\vec F_m}{\vec\eta}\vec S_{n+1-m}-
\D{\vec F_n}{\theta}\\
 & \equiv & \vec C_n-\D{\vec F_n}{\theta},
\end{eqnarray*}
where $\vec f_n$ denotes terms of order $|\vec u|^n$ in the expansion
of $\vec f(\vec x,\vec y)$ (and $\vec g_n$ is defined similarly from
the expansion of $\vec g$), and $\vec C_n$ is a multi-nomial in
$x=r\cos\theta$ and $z=r\sin\theta$ which involves quantities known
from the previous calculation of $\vec R_m$, $\vec F_m$, $\vec S_m$,
$\vec G_m$ for $m=2,\ldots,n-1$.
We choose $\vec R_n$ to be independent of $\theta$, in fact to be
$\TAV{\vec C_n}$, where the overbar denotes the mean with respect to
$\theta$, so that no secular terms arise in $\vec F_n$.
Then we choose $\vec F_n$ as simply as possible (that is, so
that $\TAV{\vec F_n}=\vec0$).  Note that $\vec R_n$ is uniquely defined
if it is to be independent of $\theta$, and if no secular terms are to
arise in $\vec F_n$.  The term $\vec F_n$ is unique only once we have
specified $\TAV{\vec F_n}$.  Altering the value of this average simply
alters the relationship between the slow co-ordinates $\vec\chi$ and
the original variables by an amount of order $|\vec u|^n$.

Now we consider the equation for the fast wave components
$\vec S_n$ and $\vec G_n$, for $n\geq2$,
\begin{eqnarray}
\vec S_n -B\vec G_n +\D{\vec G_n}{\theta} & = & 
     \vec g_n -\sum_{m=2}^{n-1}\D{\vec G_{m}}{\vec\chi}\vec R_{n+1-m}
              -\sum_{m=2}^{n-1}\D{\vec G_{m}}{\vec\eta}\vec S_{n+1-m}
\nonumber\\
& \equiv & \vec D_n.
\label{4.5}
\end{eqnarray}
We note that $\vec D_n$ can be written as a multi-nomial in $r\cos\theta$
and $r\sin\theta$ which involves quantities calculated at previous
orders.  If we now compute
$B$(\ref{4.5})${}+\partial$(\ref{4.5})$/\partial\theta$ we obtain
\begin{equation}
\left(1+\Dn{2}{ }{\theta}\right)\vec G_n=
\left(B+\D{ }{\theta}\right)(\vec D_n-\vec S_n),
\label{4.6}
\end{equation}
where we have used the identity $B^2+I=0$.  Our aim now is to make
$\dot{r}$ and $\dot{\theta}$ slowly-varying by requiring that they be
independent of $\theta$, that is,
\begin{eqnarray*}
\dot{r} & = & r\rho(\vec\chi,r^2)\\
\dot{\theta} & = & \psi(\vec\chi,r^2).
\end{eqnarray*}
But $r\dot{r}=x\dot{x}+z\dot{z}$ and $r^2\dot{\theta}=x\dot{z}-z\dot{x}$,
and so 
\[
\left[\begin{array}{cc}x & z\\-z & x\end{array}\right]\dot{\vec\eta}
=
\left[\begin{array}{cc}x & z\\-z & x\end{array}\right]
\left[\begin{array}{c}\dot{x}\\ \dot{z}\end{array}\right]
=
\left[\begin{array}{c}\rho\\ \psi\end{array}\right]r^2\,.
\]
Thus
\begin{equation}
\dot{\vec\eta}=
\left[\begin{array}{c}\dot{x}\\ \dot{z}\end{array}\right]
=
\left[\begin{array}{cc}x & -z\\
                       z & x\end{array}\right]
\left[\begin{array}{c}\rho \\ \psi \end{array}\right]
\equiv M\left[\begin{array}{c}\rho \\ \psi \end{array}\right]\,.
\label{Mrho}
\end{equation}
Jacobs~\cite[Appendix~B]{Jacobs91} has termed~(\ref{Mrho}) a ``preferred
choice'' for the form of $\dot{\vec\eta}$: at this point it is clear,
though, that~(\ref{Mrho}) is a {\em necessary} condition for $\dot{r}$
and $\dot{\theta}$ to be slowly-varying.

Equation~(\ref{Mrho}) implies that $\vec S_n$ is of the form
$\tilde{\vec S}_n(\vec\chi,r^2)\vec\eta$, where $\tilde{\vec S}_n$ is a
$2\times2$ matrix.  In~(\ref{4.6}) we choose $\vec G_n$ to match all
terms in the right-hand side, except the components of $r\cos\theta$
and $r\sin\theta$ in $\vec D_n$ (for which
$1+\partial^2/\partial\theta^2=0$), which must be removed by $\vec
S_n$.  So let us now consider these terms in~(\ref{4.5}).  We know
from~(\ref{Mrho}) that $\vec S_n$ is of the form
\[
\vec S_n=
M\left[\begin{array}{c}\rho_n(\vec\chi,r^2) \\ \psi_n(\vec\chi,r^2)
       \end{array}\right]\,.
\]
It is also straightforward to confirm that if the components of $\vec G_n$
proportional to $x=r\cos\theta$ and $z=r\sin\theta$ are
\[
\vec G_n^{(1)}=\left[ \begin{array}{c} \gamma_1x+\gamma_2z\\
                                \delta_1x+\delta_2z\\
                        \end{array}\right]\,,
\]
where the $\gamma_1$, $\gamma_2$, $\delta_1$ and $\delta_2$ are functions of 
$\vec\chi$ and $r^2$, then
\[
-B\vec G_n^{(1)}+\D{\vec G_n^{(1)}}{\theta}
=\left[\begin{array}{cc}x & z \\ -z & x 
       \end{array}\right]
\left[\begin{array}{c}\gamma_2+\delta_1\\  \delta_2-\gamma_1
      \end{array}\right]
\equiv N
\left[\begin{array}{c}\gamma_2+\delta_1\\  \delta_2-\gamma_1
      \end{array}\right]\,.
\]
We note also that the terms in $\vec D_n$ proportional to $r\cos\theta$
and $r\sin\theta$ may in all generality be written as $M\vec a+N\vec
b$, where $\vec a$, $\vec b$ are vector functions of $\vec\chi$ and
$r^2$.  This decomposition is unique, and therefore
$(\rho_n,\psi_n)=\vec a$ and
$(\gamma_2+\delta_1,\delta_2-\gamma_1)=\vec b$.  Thus we have
specified $\vec S_n=\vec a$ uniquely, although $\vec G_n$ retains two
undetermined coefficients.  The two degrees of freedom left at our
disposal correspond to redefining $r$ and $\theta$ by amounts of order
$|\vec u|^n$.

\subsection{Discussion of the normal form}

We have used the algebraic programming system {\sc reduce} to implement
the procedure described above to compute the normal form, which gives
as the first few terms of the expansion
\begin{eqnarray}
U & \sim & u-bvx+\frac{1}{4}b^2u(z^2-x^2)\nonumber\\
V & \sim & v+bux+\frac{1}{4}b^2v(z^2-x^2)\nonumber\\
W & \sim & w+bz(u^2-v^2)\label{UVWXZa}\\
X & \sim & x-buv+\frac{1}{2}b^2x(v^2-u^2)\nonumber\\
Z & \sim & z+bw(u^2-v^2).\nonumber
\end{eqnarray}
Despite the apparent differences, this normal form is equivalent to 
that of Jacobs~\cite{Jacobs91}, with his damping and forcing set to 
zero, because we have chosen our variables $\vec\chi$, $\vec\eta$ 
differently.

We now consider the structure of the normal form equations, and try to
construct their {\SIM}.  A careful examination of the co-ordinate
transformation~(\ref{4.2}) at each order $n$, approximated above, reveals
it be of the form
\begin{eqnarray}
 U & = & u + uA_1-bvA_2\nonumber\\
 V & = & v + vA_1+buA_2\nonumber\\
 W & = & w+A_3\label{UVWXZ}\\
 X & = & x+X(u,v,w)+A_4\nonumber\\
 Z & = & z+Z(u,v,w)+A_5,\nonumber
\end{eqnarray}
where $(X(u,v,w),Z(u,v,w))$ is the slow manifold $\cal M$ computed in
Section~\ref{s2}, and $A_j=A_j(u,v,w,x,z)=\Ord{r}$ as $r\rightarrow0$. 
The evolution of the slow variables is given by expressions of the form
\begin{eqnarray}
 \dot{u} & = & -vw(1-B_1(u^2,v^2,w^2,r^2))\nonumber\\
 \dot{v} & = & uw(1-B_1(v^2,u^2,w^2,r^2))\nonumber\\
 \dot{w} & = & -uv(1-r^2B_2(u^2,v^2,w^2,r^2))\label{form}\\
 \dot{r} & = & ruvwC(u^2,v^2,w^2,r^2)\nonumber\\
 \dot{\theta} & = & 1+D(u,v,w,r^2).\nonumber 
\end{eqnarray}

The equations that govern the evolution of the four variables
$(u,v,w,r)$ representing amplitudes in~(\ref{form}) are independent of
the fast gravity-wave phase variable, $\theta$.  We therefore examine
first the four-dimensional system~(\ref{form}a--d) that results from
ignoring the $\dot\theta$-equation, (\ref{form}e).  There are two
invariants of the system~(\ref{L86}), each independent of the
gravity-wave phase: $U^2+V^2$ and $V^2+W^2+X^2+Z^2$ (L86).  Similarly,
these quantities are invariant under evolution of the normal form
equations~(\ref{form}a--d), which therefore have two-dimensional
dynamics.  Solutions are therefore in general periodic.  Re-introducing
$\theta$-variations, we see that solutions of the normal
form~(\ref{form}) are in general quasiperiodic, with one frequency
$\omega_1$ arising from~(\ref{form}a--d), and a second $\omega_2$ from~(\ref{form}e).  Singly periodic solutions occur if $\omega_1$ and
$\omega_2$ are rationally related, or if $r=0$; these solutions form
the \SIM\ of the normal form, which we construct below.  The original
system~(\ref{L86}), however, not only has periodic and quasiperiodic
solutions, but also has aperiodic solutions (LK87).  Thus the normal
form and~(\ref{L86}), though quantitatively nearly identical, have
qualitatively different dynamics.

A further indication of the differences between solutions
of~(\ref{form}) and~(\ref{L86}) is the existence of some heteroclinic
orbits in the former which are absent in the latter.  To see this, we
follow LK87 in considering the Hadley solutions $P_V(F)$
of~(\ref{L86}): namely the fixed points $(U,V,W,X,Z)=(0,F,0,0,0)$.
(The same considerations will apply, with appropriate changes, to the
solutions $P_U(F)$: $(U,V,W,X,Z)=(F,0,0,0,0)$.)  Computations by LK87
indicate that gravity waves arise on the unstable manifold $U_L(F)$ of
each Hadley solution $P_V(F)$, except at isolated values of $F$.  In
these exceptional cases, $U_L(F)$ is heteroclinic to $P_V(F)$ and
$P_V(-F)$, that is, $U_L(F)$ is asymptotic to $P_V(F)$ and $P_V(-F)$ as
$t\rightarrow-\infty$ and $t\rightarrow\infty$, respectively.  The
near-heteroclinic behavior associated with almost all Hadley solutions
causes solutions of~(\ref{L86}) that pass through a neighborhood of the
Hadley solutions to be aperiodic: solutions that remain away from the
Hadley solutions are in general quasiperiodic, except on the \SIM,
where they are periodic.  However, for the system~(\ref{form}a--d) the
existence of two invariants and the symmetries of the system in any of
the $(u,v,w)$ co-ordinate planes constrain all Hadley solutions
$P_V(F)$ to be connected to their opposites $P_V(-F)$ by heteroclinic
orbits.  No aperiodic solutions of the normal form~(\ref{form}) exist.
In summary, the dynamics of~(\ref{form}) may differ qualitatively from
the dynamics of~(\ref{L86}), particularly for solutions that pass close
by the Hadley solutions.  The difference is due to the essentially
two-dimensional nature of the normal form which occurs when $\theta$ is
decoupled, compared with the essentially three-dimensional nature
of~(\ref{L86}).  The differences in the behavior of solutions
of~(\ref{L86}) and~(\ref{form}) that do not pass close to the Hadley
solutions is small.

\begin{figure} 
\centerline{\includegraphics{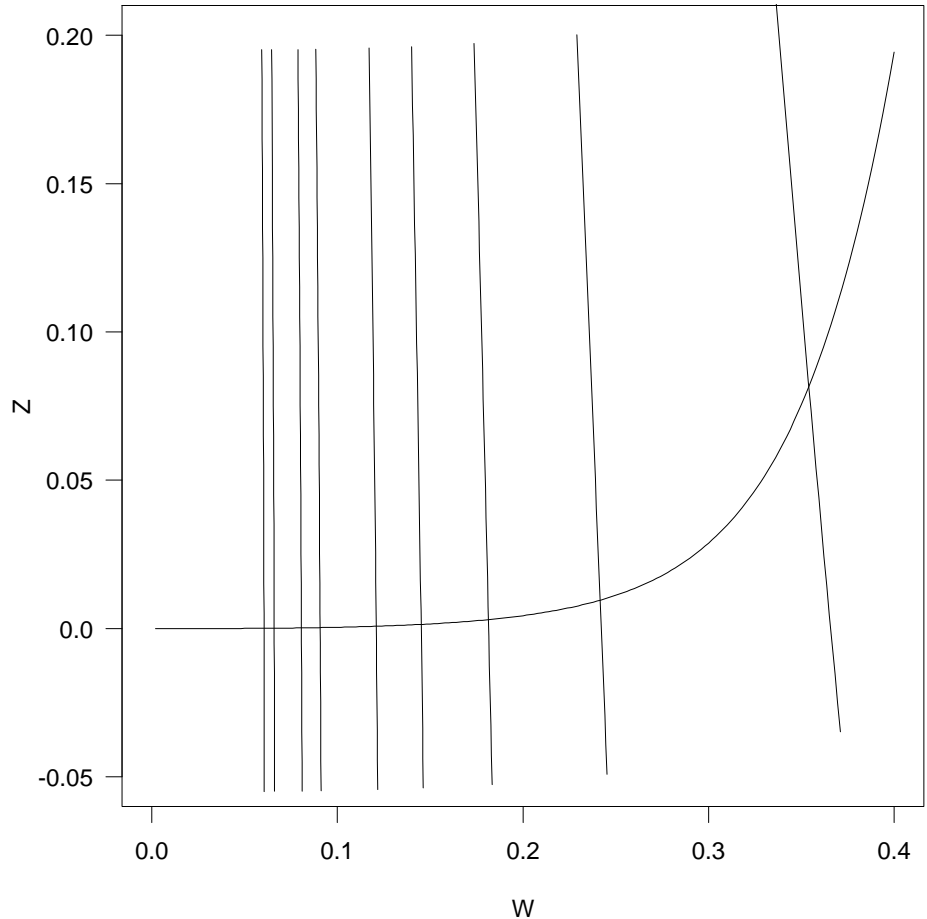}}
\caption{The slowest invariant manifold (\SIM) of the
normal form, equations~(\protect\ref{UVWXZa}) and~(\protect\ref{form})
computed to fourth order.}
\label{fignormsim} 
\end{figure}

A final illustration of the differences between solutions
of~(\ref{form}) and those of~(\ref{L86}) is given by the periodic
solutions, which form the \SIM. The \SIM\ of~(\ref{L86}) was described
in Section~\ref{s3}---to construct the \SIM\ of~(\ref{form}) we first
note that if we set $r=0$ at time $t=0$ then by~(\ref{form}d) $r=0$ for
all time, so the plane $r=0$ is an invariant slow manifold
of~(\ref{form}).  Further, solutions with $r=0$ are in general periodic
(because then the variations in $\theta$ are irrelevant when the
solution $(U,V,W,X,Z)$ is reconstructed from~(\ref{UVWXZ})).  Setting
$r=0$ in~(\ref{UVWXZ}), we see that the invariant manifold $r=0$ is
precisely $\cal M$, as calculated in Section~\ref{s2}.  Some other
periodic solutions of~(\ref{form}) occur for non-zero initial values of
$r$---however, these have significant gravity-wave activity and they
exist when the gravity waves happen to be exact harmonics of the Rossby
waves.  They give rise to manifolds of resonant solutions that
intersect $\cal M$, akin to the lines $W^*=K/(k\pi)$ for the uncoupled
system~(\ref{L86}) with $b=0$.  The slowest invariant manifold of the
normal form~(\ref{form}) is therefore $\cal M$, and has no
singularities.  However, there are infinitely many ``resonant''
branches, which intersect $\cal M$ as shown in Figure~\ref{fignormsim}.


The relation between the normal form calculation of this section and
the calculation of the slow manifold described in Section~\ref{s2} is most
easily seen if we assume that the new variables $\vec\chi$ are chosen,
as we have done, by setting $\TAV{\vec F}=\vec0$.  Then the
transformation of the Rossby wave amplitudes is of the form
\[
\vec x=\vec\chi+\vec F(\vec\chi,\vec\eta)
=\vec\chi+r\tilde{\vec F}(\vec\chi,\vec\eta),
\]
so that by setting $r=0$ (that is, $\vec F=\vec\eta=\vec0$) the new slow
variables and the old become identical.  To see the significance of
setting $r=0$ we consider~(\ref{4.3a}--\ref{4.3b}), which become
\begin{eqnarray*}
\dot{\vec x} & = & \vec f(\vec x,\vec h(\vec x))\\
\D{\vec h}{\vec x}\dot{\vec x} & = & B\vec h(\vec x)+\vec g(\vec x),
\end{eqnarray*}
where $\vec h(\vec x)=\vec G(\vec x,\vec0)$.  These are precisely the
equations that govern $\dot{\vec x}$ and $\vec y=\vec h(\vec x)$ for
an invariant slow manifold (Carr~\cite{Carr81}).  So by setting
$\vec\eta=\vec0$ in~(\ref{4.2}) we recover $\cal M$ (implicitly,
through $\vec y=\vec h(\vec x)$).  As a consequence, we expect the
normal form sums in~(\ref{FG}) and~(\ref{RS}) to be divergent.


There is a strong parallel between the method of normal forms and the
method of averaging. (The latter method was suggested for the
geostrophic approximation by Bill Dewar in private communication.)  An
alternative approach to the approximation of solutions of~(\ref{L86})
is to assume that the gravity wave oscillations occur on a separate,
faster time-scale than the evolution of the other variables, which are
assumed to vary on the slow time-scale $\tau$.  We then replace the
operator $d/dt$ in~(\ref{L86}) by the operator $d/d\tau+d/d\theta$, and
write each dependent variable in~(\ref{L86}) as a function of the new
variables $u(\tau)$, $v(\tau)$, $w(\tau)$, $a(\tau)$ and the fast time
$\theta$.  The notation for the new variables is as for the normal form
calculation, except that the gravity waves now have a slowly-varying
complex amplitude $a(\tau)$.  The two-time approach proceeds with
similar algebraic steps as the normal form calculation, in particular at
each stage an average over the fast time-scale $\theta$ is taken in
order to find  terms in the slow evolution equations.

\subsection{Projection of initial conditions and the normal form
transformation}


If the slow manifold had been a centre manifold,\footnote{This
    is the case when damping of the gravity waves is incorporated
    into the model (Cox \& Roberts~\cite{Cox91}).} characterized by an
exponential approach of nearby trajectories to $\cal M$, then a
rational principle can be used  to compute the balanced initial data
(Roberts~\cite{Roberts89b}).  The principle is that the long-term evolution of the
original full system from its initial data must approach exponentially
quickly the slow evolution on the centre manifold from the balanced
initial data. This principle leads to an analysis whose framework is
identical to that given in Subsection~\ref{s2a}. However, the balanced initial
data can be found immediately from the normal form of the
equations, as we now demonstrate. Continuing the discussion for the
case when $\cal M$ is a centre manifold, we find that a normal
form transformation~(\ref{4.2}) can be chosen so that the new variables 
evolve according to equations of the form
\begin{eqnarray}
\dot{\vec\chi}&=&\vec R(\vec\chi),\label{cmnormx}\\
\dot{\vec\eta}&=&B\vec\eta+\vec S(\vec\chi,\vec\eta)
               =\left(B+ \tilde{\vec S}(\vec\chi,r^2) \right)\vec\eta.
\label{cmnormy}
\end{eqnarray}
Thus $\vec\eta=\vec0$ describes the centre manifold of the
long-term evolution and, because the eigenvalues of $B$ are
negative, $\vec\eta$ decays exponentially quickly to $\vec0$ (as
is the case when the fast gravity waves are damped).  Thus~(\ref{4.2}) gives
the following parametric description of the centre manifold in the
original variables
\[
\vec x=\vec\chi+\vec F(\vec\chi,\vec0),\qquad
\vec y=\vec G(\vec\chi,\vec0).
\]
However, from equation~(\ref{cmnormx}) we see that the evolution of 
the slow variables $\vec\chi$ is unaffected by the fast variables 
$\vec\eta$.  Thus all solutions starting from points 
$(\vec\chi,\vec\eta)$ with the same value of $\vec\chi$ have precisely 
the same $\vec\chi$ evolution, and because they all asymptote 
exponentially quickly to the centre manifold, so they all have the 
same long-term evolution.  The hyper-surfaces in $(\vec x,\vec 
y)$-space that are described as $\vec\eta$ is varied in~(\ref{4.2}) 
with fixed $\vec\chi$ form the {\em isochronic manifolds} introduced 
by Roberts~\cite{Roberts89b}.  Hence, the initial data 
$(\vec\chi,\vec\eta)$ are balanced to the centre manifold simply by 
setting $\vec\eta=\vec0$ (as in a linear analysis).  In terms of the 
original variables, the initial values $(\vec x_0,\vec y_0)$ are 
balanced by finding the corresponding normal form variables 
$(\vec\chi_0,\vec\eta_0)$ from~(\ref{4.2}), and then using
\[
{\vec x}'_0=\vec\chi_0+\vec F(\vec\chi_0,\vec0),\qquad
{\vec y}'_0=\vec G(\vec\chi_0,\vec0)
\]
as the appropriate initial conditions.  From $(\vec x'_0,\vec y'_0)$ the
evolution is slow and has precisely the same long-term dynamics as from
$(\vec x_0,\vec y_0)$.

However, the ``quasi-geostrophic'' slow manifold $\cal M$ is a {\em
subcentre} manifold; it does not attract neighboring solutions
exponentially, and the arguments above do not directly apply.  The
reason is that for a subcentre manifold the influence of the fast
 waves cannot be removed entirely from the evolution of the slow
variables.   This may be seen in~(\ref{form}) where the functions $B_1$
and $B_2$, which govern the evolution of the slow  waves, also
depend upon the amplitude of the fast  waves, $r$:  for a
subcentre manifold~(\ref{cmnormx}) must be replaced, in general, by 
an equation of the form
\[
\dot{\vec\chi}=\vec R(\vec\chi,r^2).
\]
That is, the evolution of the slow variables cannot be entirely decoupled
from the fast variables.  Thus a simulation with both Rossby waves
{\em and} gravity waves present need not be equivalent, over a long
time, to any of the possible solutions with purely slow Rossby waves.

This feature of subcentre manifolds is displayed in the simple system
\[
\dot u=z^2,\qquad \dot x=-z,\qquad \dot z=x.
\]
These equations have the exact normal form,
upon the transformation $\chi=u+xz/2$, $x=r\cos\theta$ and $z=r\sin\theta$,
\[
\dot\chi=r^2/2,\qquad \dot r=0,\qquad \dot \theta=1.
\]
Here the slow manifold is simply $x=z=0$ ($r=0$), on which $\chi$ evolves according
to $\dot\chi=0$. Thus the long-term evolution on $\cal M$ is trivial: $\chi$
is constant.  However, if there are any fast waves present, $r\neq0$,
then $\chi$ drifts according to $\dot \chi=r^2/2$, but there is no
corresponding solution for $\chi$ on $\cal M$.  The same is true for the
normal form~(\ref{form}) of the model L86---at fourth order in
$|\vec\chi|$ we cannot avoid introducing gravity-wave terms proportional to
$r^2$ into the evolution equations for the slow Rossby waves.  Therefore
solutions on and off the slow manifold $\cal M$ cannot share the same
values of $u$, $v$ and $w$ for all time; there are inevitable
discrepancies which become significant on a timescale of $\Ord{r^{-2}}$.

It is now clear that the balancing procedure, the projection of 
initial conditions described in Section~\ref{s2}, which is linear in 
the fast gravity wave amplitude $r$, cannot be improved to be correct 
to $\Ord{r^2}$.

\section{Conclusions}
\label{s5}

A significant part of our understanding of atmospheric dynamics rests 
on the concepts of quasi-geostrophy (Gill~\cite[Chapt.~7]{Gill82}).  
The fundamental concept is the separation of the dynamics into waves 
of two time-scales: the slowly evolving Rossby waves and the quickly 
evolving gravity waves.  We have described how a slow manifold, 
formally composed of the ensemble of slow Rossby waves, may formally 
be written in the form $\vec y=\vec h(\vec x)$, with $\vec h$ 
developed as a power series in the slow variables $\vec x$.  This is 
similar to the scheme proposed by Baer \& Tribbia~\cite{Baer77}, and 
is equivalent to that of Vautard \& Legras~\cite{Vautard86}.  As L86 
demonstrates, the power series for $\vec h$ is divergent, so that 
successive approximation schemes at first appear to converge to a slow 
manifold, but after the inclusion of sufficiently many terms they 
begin to diverge.  Nevertheless, the divergence of its power series 
does not indicate the non-existence of a slow manifold.  Each 
successive approximation to $\cal M$ captures the dynamics of L86 to 
within a higher power of $\vec x$, but in a smaller radius around the 
origin.  Unfortunately for our purposes, the dynamics on $\cal M$ 
(determined exactly, but not from its asymptotic series) differ from 
those of L86 by an amount that is smaller than any power of $\vec x$ 
as $\vec x\rightarrow\vec0$.  The dynamics of the closed set of 
evolution equations~(\ref{xdot}) for the slow variables on $\cal M$ 
are genuinely slow: no gravity waves develop.  But $\cal M$ is not 
quite invariant, so the apparent slow dynamics on $\cal M$ do not 
reflect genuine slow dynamics in L86.  In particular $\cal M$ appears 
to be filled with periodic solutions, whereas the full system has a 
manifold of periodic solutions (Lorenz' {\SIM}) which is different 
from $\cal M$ and which also contains singularities.

The issue of how to balance given data measurements by projecting them 
onto a slow manifold is a complicated one for realistic 
situations---the most appropriate projection may depend on the 
relative quality of the various measurements, and on the influence of 
different measurements on the projected initial point (Daley~\cite{Daley80,Daley81}).  We have described a method of projection 
that is certainly well-suited to models where the initial data are 
known exactly, such as L86.  The criterion we have proposed for 
selecting an initial point on $\cal M$ seems appropriate for numerical 
weather prediction: namely, that the behavior of the balanced system 
should correspond for all time to that of the unbalanced system (just 
without the fast gravity-wave activity).  This ensures that the 
essential features of the forecast are unchanged by the initialization 
process.  However, we have also shown that due to resonances, as 
exhibited by the normal form transformation (Arnold~\cite{Arnold82}), the 
initialization can be carried out only to first order in the fast wave 
amplitude\footnote{ The resonances that force certain terms to occur 
in the evolution equations for the normal form variables are a 
consequence of the {\em linear} dynamics (Arnold~\cite{Arnold82}), and are quite 
distinct to the resonances between the periods of the (fully 
nonlinear) Rossby and gravity waves that induce singularities in the 
slowest invariant manifold.}.  We have illustrated the practical 
utility of our proposed initialization procedure with numerical 
integrations (Figure~\ref{ic}) that show the superior accuracy of the 
forecast from appropriately initialized data.

Certain difficulties associated with computing a slow
manifold have led researchers to question the existence of such a
manifold, or to label the concept ``fuzzy''.  A slow manifold for L86
is a subcentre manifold (Kelley~\cite{Kelley67}; Sijbrand~\cite{Sijbrand85}), and is not
unique: just as for all low-dimensional models (Roberts~\cite{Roberts89}) there
are infinitely many slow manifolds in which the fast variables are specified
in terms of the slow variables. It is this non-uniqueness that accounts
for the ``fuzziness'' in the concept of the slow manifold: the partial
differential equation that governs a slow manifold for a given problem
is perfectly well-defined, as is each manifold. It is just that the
construction of low-dimensional models has infinitely many solutions,
differing by exponentially small terms.  It is possible to select one
distinguished slow manifold by assuming that the fast variables may be
written as power series in the slow variables.  However, often, as is
the case for L86, the power series for the slow manifold proves to be
divergent, so that the series is asymptotic.  As an alternative, the
numerical construction of a slow manifold (Lorenz' {\SIM}) was
conceived to avoid the divergence problem.  The {\SIM} is defined as
containing all the periodic solutions.  However, in the light of our
previous comment that slow manifolds differ by exponentially small
terms, the {\SIM} must share the divergent power series of all slow
manifolds for L86.  This does not imply that the {\SIM} does not exist,
nor does it suggest that the {\SIM} is somehow ``fuzzy''.  The {\SIM}
exists; it just has a divergent power series.  The asymptotic series of
all slow manifolds for a given system are identical---the slow
manifolds differ by {\em sub-dominant terms} (Bender and Orszag~\cite{Bender81}).

We contrast our point of view with that presented by Warn \& Menard~\cite{Warn86}, relating to the slow manifold for the model L80.  They argue
that a slow manifold must attract neighboring solutions in order to be
useful in applications.  Since almost every numerical solution of L80
involves some level of fast ``gravity'' wave activity, they conclude that an
attracting invariant slow manifold does not exist.  However, it is
important to recognize that a slow manifold need not be attracting in
order to serve as a useful centre for the dynamics of the system.  The
subcentre manifold we have discussed for L86 does not attract neighboring
solutions, yet it has dynamics that approximately correspond to the full
initial-value problem, except that the fast gravity waves are absent. Warn and
Menard argue that the slow manifold should be replaced by a more
general ``fuzzy'' balanced set because their scheme to compute a slow
manifold encounters asymptotic series, and because in their
approximations to the slow manifold, gravity wave activity is
observed.  We have argued that a slow manifold is necessarily ``fuzzy''
because of its exponential non-uniqueness, which is inherent to
invariant manifolds of the centre-manifold family.  Each individual
invariant manifold is well-defined.  The ``fuzziness'' of the set of
manifolds is an entirely separate issue from that of the divergence of
the power series for each slow manifold.  As Lorenz has illustrated for
the {\SIM}, one may compute a slow manifold numerically, even
when it has a divergent series.

We cite two other examples of common and useful approximations that 
correspond to the use of a subcentre manifold, even though such a 
manifold does not attract neighboring solutions.  The first is the 
incompressible approximation in fluid mechanics, where the fast 
variables represent sound waves and are neglected to leave the slow 
evolution of incompressible flow.  The second is beam theory in 
elasticity, where rapid internal elastic vibrations are the fast 
variables and where large scale deformations are the slow variables 
(Roberts~\cite{Roberts93}).  The fast ``ringing'' modes are eliminated in the 
traditional approximations of beam theory.  The concept of a subcentre 
manifold enables one rationally to extend these traditional 
approximations to higher order, to derive appropriate initial and 
boundary conditions for the model, and to treat forcing appropriately.

\newpage
\appendix

\section{Contour integration}

In order to compute $Z_1^*$, we wish to evaluate the integral
\[
I=U^{*2}\int^{T_0}_0 \sin (T_0-\tau)\,\CN(W^*\tau)\,\SN(W^*\tau)\,d\tau,
\]
which appears in~(\ref{goApp}).
We first expand the trigonometric function, then change variables by 
setting $t=W^*\tau$, so that
\begin{equation}
I=mW^*\left(I_c \sin T_0 - I_s \cos T_0\right),
\label{IcIs}
\end{equation}
where 
\[
I_c=\int_0^K\cos\frac{t}{W^*}\,\SN(t)\,\CN(t)\,dt,
\]
and
\[
I_s=\int_0^K\sin\frac{t}{W^*}\,\SN(t)\,\CN(t)\,dt.
\]

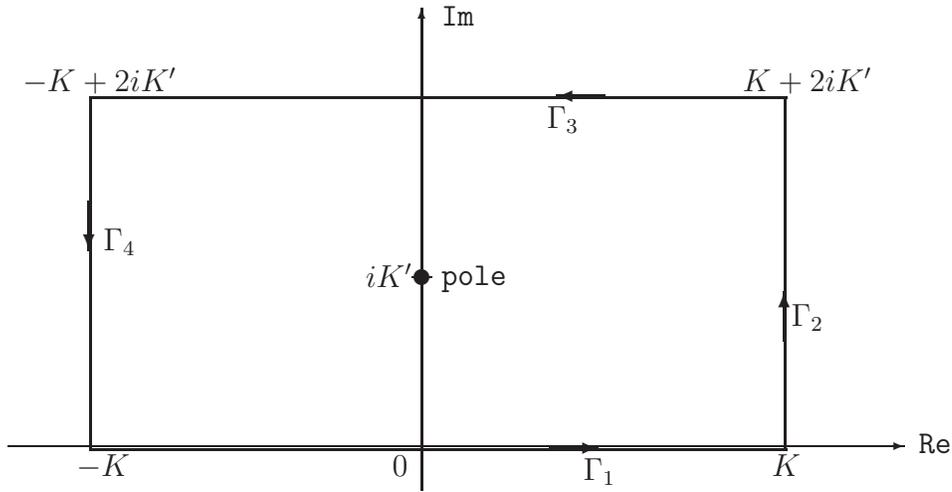
\begin{figure}
\begin{center}
{\tt    \setlength{\unitlength}{0.075em}
\begin{picture}(410,218)
\thinlines    \put(49,108){$\Gamma_4$}
              \put(245,14){$\Gamma_1$}
              \put(230,158){$\Gamma_3$}
              \put(330,77){$\Gamma_2$}
              \put(179,97){\circle*{6}}
              \put(187,94){pole}
              \put(167,16){$0$}
              \put(156,93){$iK'$}
              \put(175,97){\line(1,0){8}}
              \put(16,173){$-K+2iK'$}
              \put(310,173){$K+2iK'$}
              \put(38,16){$-K$}
              \put(322,16){$K$}
 \thicklines
              \put(43,128){\vector(0,-1){20}}
              \put(254,171){\vector(-1,0){20}}
              \put(327,71){\vector(0,1){20}}
              \put(231,27){\vector(1,0){20}}
\thinlines    \put(382,25){Re}
              \put(187,200){Im}
              \put(179,10){\vector(0,1){197}}
              \put(10,28){\vector(1,0){366}}
\thicklines   \put(44,27){\framebox(283,143){}}
\end{picture}}
\end{center}
\caption{Paths of integration in the complex plane.}
\label{figint}
\end{figure}

To evaluate $I_s$, we consider the contour integral
\begin{equation}
I_\Gamma=\int_\Gamma S(t;W^*)\,dt,
\label{A1}
\end{equation}
where
\[
S(t;W^*)= e^{it/W^*}\SN(t)\,\CN(t),
\]
and where $\Gamma$ is the rectangle with vertices at $-K$, $K$, $K+2iK'$,
$-K+2iK'$ as shown in Figure~\ref{figint}.
We denote the straight line segments joining these vertices from $-K$
counterclockwise by $\Gamma_1$, $\Gamma_2$, $\Gamma_3$, $\Gamma_4$,
respectively, with corresponding integrals $I_1$, $I_2$, $I_3$, $I_4$.
Then
\[
I_1=\int_{-K}^K e^{it/W^*}\SN(t)\,\CN(t)\,dt=2iI_s.
\]
The integral along the parallel side of $\Gamma$ is
\begin{eqnarray*}
I_3 & = & \int^{-K}_K e^{i(2iK'+t)/W^*}\SN(2iK'+t)\,\CN(2iK'+t)\,dt\\
    & = & e^{-2K'/W^*}I_1.
\end{eqnarray*}
The other integrals along the vertical sides of the rectangle give
\begin{eqnarray*}
I_2 & = & \int_0^{2K'} e^{i(K+it)/W^*}\SN(K+it)\,\CN(K+it)\,idt\\
    & = & m_1^{1/2}e^{iK/W^*}\int_0^{2K'}e^{-t/W^*} \ND(t|m_1)\,\SD(t|m_1)\,dt,
\end{eqnarray*}
where $m_1=1-m$, and
\[
I_4=-I_2^*.
\]
The contour $\Gamma$ encloses a double pole of $S(t;W^*)$ at $t=iK'$, and  so
\[
I_\Gamma = I_2-I_2^*+2i(1+e^{-2K'/W^*})I_s =
2\pi i \,\mbox{Residue}(S(t;W^*);t=iK').
\]
To compute the residue we expand $S(t;W^*)$ near the pole to give
\[
S(t;W^*)=e^{-K'/W^*}\left(1+i\frac{t-iK'}{W^*}\right)
                    \left(\frac{m^{-1/2}}{t-iK'}\right)
                    \left(\frac{-im^{-1/2}}{t-iK'}\right)+
\Ord{1},
\]
where the singular behavior of the Jacobian elliptic functions
is (Table~16.7 of Abramowitz and Stegun~\cite{Abramowitz64})
\[
\SN(t)\sim\frac{m^{-1/2}}{t-iK'},\quad
\CN(t)\sim\frac{-im^{-1/2}}{t-iK'}\quad\mbox{ as }t\rightarrow iK'.
\]
Thus
\[
\mbox{Residue}(S(t;W^*);t=iK')=\frac{1}{mW^*}e^{-K'/W^*}.
\]
Therefore 
\[
I_s=\frac{\pi}{2mW^*}\sech\frac{K'}{W^*}-
\frac{m_1^{1/2}}{1+e^{-2K'/W^*}}
\int_0^{2K'}e^{-t/W^*}\,\ND(t|m_1)\,\SD(t|m_1)\,dt \sin T_0.
\]
Through successive integrations by parts we find that the integral in 
this expression has the asymptotic series
\begin{eqnarray*}
\int_0^{2K'}e^{-t/W^*}\,\ND(t)\,\SD(t)\,dt  & = & 
\int_0^{2K'}e^{-t/W^*}\,\left(-\frac{1}{m}\frac{d}{dt}\CD(t)\right)\,dt\\
 & \sim & -\frac{1}{m}\left(1+e^{-2K'/W^*}\right)
\sum_{n=1}^\infty W^{*2n}\CD^{(2n)}(0).
\end{eqnarray*}
Thus we may write
\[
I_s\sim \frac{\pi}{2mW^*}\sech\frac{K'}{W^*}+
\frac{m_1^{1/2}}{m} \sum_{n=1}^\infty W^{*2n}\CD^{(2n)}(0) \sin T_0.
\]

Now we note that the integral $I_c$ has the asymptotic series given by
\begin{eqnarray*}
I_c & = & \int_0^K\cos\frac{t}{W^*}\,\SN(t)\,\CN(t)\,dt\\
& = &
\int_0^K\cos\frac{t}{W^*}\,\left(-\frac{1}{m}\frac{d}{dt}\DN(t)\right)\,dt\\
 & \sim & 
-\frac{1}{m}\sum_{n=1}^\infty W^{*2n}(-1)^{n-1}\left(
\DN^{(2n)}(K)\cos T_0 -\DN^{(2n)}(0) \right)
\end{eqnarray*}
and that
\[
\DN^{(2n)}(K)=m_1^{1/2}\ND^{(2n)}(0)=(-1)^nm_1^{1/2}\CD^{(2n)}(0)
\]
because $\ND(it|m)=\CD(t|m_1)$.
Therefore, by substituting the expressions we have calculated for $I_s$
and $I_c$ into~(\ref{IcIs}), we find
\begin{eqnarray*}
I & \sim & mW^*\left(
\left[
\frac{m_1^{1/2}}{m}\sum_{n=1}^\infty W^{*2n} \CD^{(2n)}(0)\cos T_0
-\frac{1}{m}\sum_{n=1}^\infty W^{*2n} \DC^{(2n)}(0)
\right]
\sin T_0
\right. \\
 & & \left. {}-
\left[
\frac{\pi}{2mW^*}\sech\frac{K'}{W^*}+
\frac{m_1^{1/2}}{m} \sum_{n=1}^\infty W^{*2n}\CD^{(2n)}(0) \sin T_0
\right]
\cos T_0
\right)\\
 & \sim & 
W^*\left(
\left[ -\frac{1}{m}\sum_{n=1}^\infty W^{*2n} \DC^{(2n)}(0) \right] \sin T_0
- \frac{\pi}{2mW^*}\sech\frac{K'}{W^*} \cos T_0
\right).
\end{eqnarray*}

This is the expression we desire for $I$. The sum that appears is divergent,
because the Taylor expansion
\[
\DC(W^*)=\sum_{n=0}^\infty \frac{1}{(2n)!} W^{*2n} \DC^{(2n)}(0)
\]
has a finite radius of convergence (equal to $K$).
It follows now from~(\ref{goApp}) that 
\begin{eqnarray*}
Z_1^* & = & -I/\sin T_0\\
      & = & \sum_{n=1}^\infty W^{*2n+1}\DC^{(2n)}(0)+
\frac{\pi}{2}\sech\frac{K'}{W^*}\cot T_0.
\end{eqnarray*}

\bibliographystyle{plain}\bibliography{ajr,bib,new}

\begin{thebibliography}{10}

\bibitem{Abramowitz64}
M.~Abramowitz and I.A. Stegun, editors.
\newblock {\em Handbook of mathematical functions}.
\newblock Dover, 1965.

\bibitem{Arnold82}
V.I. Arnold.
\newblock {\em Geometrical methods in the theory of ordinary differential
  equations}.
\newblock Springer, 1982.

\bibitem{Baer77}
F.~Baer and J.J. Tribbia.
\newblock On complete filtering of gravity modes through nonlinear
  initialization.
\newblock {\em Mon. Wea. Rev.}, 105:1536--1539, 1977.

\bibitem{Bender81}
C.M. Bender and S.A. Orszag.
\newblock {\em Advanced mathematical methods for scientists and engineers}.
\newblock McGraw-Hill, 1981.

\bibitem{Camassa95}
R.~Camassa.
\newblock On the geometry of an atmospheric slow manifold.
\newblock {\em Physica D}, 84:357--397, 1995.

\bibitem{Carr81}
J.~Carr.
\newblock {\em Applications of centre manifold theory}, volume~35 of {\em
  Applied Math Sci}.
\newblock Springer-Verlag, 1981.

\bibitem{Cox91}
S.M. Cox and A.J. Roberts.
\newblock Centre manifolds of forced dynamical systems.
\newblock {\em J. Austral. Math. Soc. B}, 32:401--436, 1991.

\bibitem{Cuyt84}
A.~Cuyt.
\newblock Pade approximants for operators: theory and application.
\newblock {\em Lect. Notes Mathematics}, 1065, 1984.

\bibitem{Daley80}
R.~Daley.
\newblock On the optimal specification of the initial state for deterministic
  forecasting.
\newblock {\em Mon. Wea. Rev.}, 108:1719--1735, 1980.

\bibitem{Daley81}
R.~Daley.
\newblock Normal mode initialization.
\newblock {\em Rev. Geophys. Space Phys.}, 19:450--468, 1981.

\bibitem{Errico91}
R.M. Errico.
\newblock Theory and application of nonlinear normal mode initialization.
\newblock Technical report, NCAR note TN--344$+$IA, 1991.

\bibitem{Gill82}
A.E. Gill.
\newblock {\em Atmosphere-Ocean Dynamics}.
\newblock Academic Press, 1982.

\bibitem{Grad63}
H.~Grad.
\newblock Asymptotic theory of the {Boltzmann} equation.
\newblock {\em Phys. Fluids}, 6:147--181, 1963.

\bibitem{Guckenheimer83}
J.~Guckenheimer and P.~Holmes.
\newblock {\em Nonlinear oscillations, dynamical systems, and bifurcations of
  vector fields}.
\newblock Springer-Verlag, 1983.

\bibitem{Haake83}
F.~Haake and M.~Lewenstein.
\newblock Adiabatic drag and intial slip in random processes.
\newblock {\em Phys. Rev.~A}, 28:3060--3612, 1983.

\bibitem{Houghton89}
J.T. Houghton.
\newblock {\em The physics of atmospheres}.
\newblock CUP, 2nd edition, 1989.

\bibitem{Jacobs91}
S.J. Jacobs.
\newblock Existence of a slow manifold on a model system of equations.
\newblock {\em J.~Atmos. Sci.}, 48:893--901, 1991.

\bibitem{Kelley67}
A.~Kelley.
\newblock On the {Liapunov} subcenter manifold.
\newblock {\em J.~Math. Anal. Appl.}, 18:472--478, 1967.

\bibitem{Leith80}
C.E. Leith.
\newblock Nonlinear normal mode initialisation and quasi-geostrophic theory.
\newblock {\em J.~Atmos. Sci.}, 37:958--968, 1980.

\bibitem{Lorenz87}
E.~Lorenz and Krishnamurty.
\newblock On the non-existence of a slow manifold.
\newblock {\em J.~Atmos. Sci.}, 44:2940--2950, 1987.

\bibitem{Lorenz80}
E.N. Lorenz.
\newblock Attractor sets and quasi-geostophic equilibrium.
\newblock {\em J.~Atmos. Sci.}, 37:1685--1699, 1980.

\bibitem{Lorenz86}
E.N. Lorenz.
\newblock On the existence of a slow manifold.
\newblock {\em J.~Atmos. Sci.}, 43:1547--1557, 1986.

\bibitem{Mercer90}
G.N. Mercer and A.J. Roberts.
\newblock A centre manifold description of contaminant dispersion in channels
  with varying flow properties.
\newblock {\em SIAM J. Appl. Math.}, 50:1547--1565, 1990.

\bibitem{Mercer94a}
G.N. Mercer and A.J. Roberts.
\newblock A complete model of shear dispersion in pipes.
\newblock {\em Jap. J. Indust. Appl. Math.}, 11:499--521, 1994.

\bibitem{Roberts85b}
A.J. Roberts.
\newblock Simple examples of the derivation of amplitude equations for systems
  of equations possessing bifurcations.
\newblock {\em J.~Austral. Math. Soc. B}, 27:48--65, 1985.

\bibitem{Roberts89b}
A.J. Roberts.
\newblock Appropriate initial conditions for asymptotic descriptions of the
  long term evolution of dynamical systems.
\newblock {\em J. Austral. Math. Soc. B}, 31:48--75, 1989.

\bibitem{Roberts89}
A.J. Roberts.
\newblock The utility of an invariant manifold description of the evolution of
  a dynamical system.
\newblock {\em SIAM J. Math. Anal.}, 20:1447--1458, 1989.

\bibitem{Roberts93}
A.J. Roberts.
\newblock The invariant manifold of beam deformations. part 1:the simple
  circular rod.
\newblock {\em J.~Elas.}, 30:1--54, 1993.

\bibitem{Sijbrand85}
J.~Sijbrand.
\newblock Properties of centre manifolds.
\newblock {\em Trans. Amer. Math. Soc.}, 289:431--469, 1985.

\bibitem{vanKampen85}
N.G. van Kampen.
\newblock Elimination of fast variables.
\newblock {\em Physics Reports}, 124:69--160, 1985.

\bibitem{Vandyke84}
M.~Vandyke.
\newblock Computer-extended series.
\newblock {\em Annu. Rev. Fluid Mech.}, 16:287--310, 1984.

\bibitem{Vautard86}
R.~Vautard and B.~Legras.
\newblock Invariant manifolds, quasi-geostrophy and initialisation.
\newblock {\em J.~Atmos. Sci.}, 43:565--584, 1986.

\bibitem{Warn86}
T.~Warn and R.~Menard.
\newblock Nonlinear balance and gravity-inertial saturation in a simple
  atmospheric model.
\newblock {\em Tellus}, 38A:285--294, 1986.

\bibitem{Watt95}
S.D. Watt, A.J. Roberts, and R.O. Weber.
\newblock Dimensional reduction of a bushfire model.
\newblock {\em Mathl. Comput. Modelling}, 21(9):79--83, 1995.

\end{thebibliography}

\end{document}